\documentclass[12pt]{article}
\pdfoutput=1
\usepackage{putex}
\usepackage{feyn}
\usepackage[vcentermath]{youngtab}
\usepackage{subfig}
\usepackage{lscape}

\usepackage{graphicx}
\usepackage{epstopdf}
\usepackage{enumerate}
\usepackage{cite}
\usepackage{tensor}
\usepackage{slashed}
\usepackage{amsmath}
\usepackage{amssymb}
\usepackage{mathrsfs}
\usepackage{lgrind}

\usepackage{bbm}

\usepackage{hyperref}

\numberwithin{equation}{section}

\newcommand {\be} {\begin {equation}}
\newcommand {\ee} {\end {equation}}

\newcommand {\bes} {\begin {equation*}}
\newcommand {\ees} {\end {equation*}}

\newcommand{\eps}{\epsilon}

\def\CH{{\cal H}}

\newcommand{\beq}{\begin{equation}}
\newcommand{\eeq}{\end{equation}}

\def\be{ \begin{equation} }
\def\ee{ \end{equation} }

\def \al{\alpha}
\def \bchi{\bar{\chi}}
\begin{document}

\preprint{PUPT-2531}

\institution{PU}{Department of Physics, Princeton University, Princeton, NJ 08544}
\institution{PCTS}{Princeton Center for Theoretical Science, Princeton University, Princeton, NJ 08544}

\title{ Spectra of Operators in Large $N$ Tensor Models
}

\authors{Ksenia Bulycheva,\worksat{\PU} Igor R.~Klebanov\worksat{\PU,\PCTS}, Alexey Milekhin\worksat{\PU}, \\[10pt] 
Grigory Tarnopolsky\worksat{\PU}}

\abstract{We study the operators in the large $N$ tensor models, focusing mostly on the fermionic quantum mechanics with $O(N)^3$ symmetry which may be
either global or gauged. In the model with global symmetry we study the spectra
of bilinear operators, which are in either the symmetric traceless or the antisymmetric representation of one of the $O(N)$ groups. In the symmetric traceless case, the spectrum
of scaling dimensions is the same as in the SYK model with real fermions; it includes the $h=2$ zero-mode. For the operators anti-symmetric in the two indices, the scaling dimensions are the same as in the additional sector found in the complex tensor and
SYK models; the lowest $h=0$ eigenvalue corresponds to
the conserved $O(N)$ charges. A class of singlet operators may be constructed from contracted combinations of $m$ symmetric traceless or antisymmetric two-particle operators.
Their two-point functions receive contributions from $m$ melonic ladders. Such multiple ladders are a new phenomenon in the tensor model, which does not seem
to be present in the SYK model. 
The more typical $2k$-particle operators do not receive any ladder corrections and have quantized large $N$ scaling dimensions $k/2$. 
We construct pictorial representations of various singlet operators with low $k$. For larger $k$ we use available techniques to count the operators and show that their 
number grows as $2^k k!$. As a consequence, the theory has a Hagedorn phase
transition at the temperature which approaches zero in the large $N$ limit.
We also study the large $N$ spectrum of low-lying operators in the Gurau-Witten model, which has $O(N)^6$ symmetry. We argue that it corresponds to one of the generalized SYK models 
constructed by Gross and Rosenhaus. Our paper also includes studies of the invariants in large $N$ tensor integrals with various symmetries.
}

\date{}
\maketitle

\tableofcontents

\section{Introduction and Summary}

Models where the degrees of freedom are tensors of rank $r>2$ offer the possibility of large $N$ limits dominated by the so-called melon diagrams, if
the interactions are chosen appropriately
\cite{Gurau:2009tw,Gurau:2011xp,Gurau:2011aq,Gurau:2011xq,Bonzom:2011zz,Tanasa:2011ur,Bonzom:2012hw,Dartois:2013he,Tanasa:2015uhr,Carrozza:2015adg,Gurau:2016lzk}.
In models where the tensor indices are distinguishable, so that the symmetry group is $O(N)^r$ for example, the proofs of melonic
limits have been available for several years.\footnote{
There is recent evidence \cite{Klebanov:2017nlk,Gurau:2017qya} that
the melon dominance extends even to theories with a single $O(N)$ symmetry group, which are similar to the tensor models \cite{Ambjorn:1990ge,Sasakura:1990fs,Gross:1991hx}
considered in the early 90s.}
During the recent months, interest in the melonic large $N$ tensor models has been boosted by their connections \cite{Witten:2016iux,Klebanov:2016xxf} with the 
Sachdev-Ye-Kitaev model \cite{Sachdev:1992fk,1999PhRvB..59.5341P, 2000PhRvL..85..840G,Kitaev:2015}
and its generalizations \cite{Gross:2016kjj}, as well as by connections with the large $N$ matrix models \cite{Ferrari:2017ryl}. 
In particular, the Schwinger-Dyson equations which determine the scaling dimensions of a class of bilinear operators
\cite{Kitaev:2015,Polchinski:2016xgd,Maldacena:2016hyu,Jevicki:2016bwu,Gross:2016kjj}
have been shown to be identical in the tensor and SYK models \cite{Klebanov:2016xxf}.

In this paper we continue exploration of the large $N$ tensor models, in particular the $O(N)^3$ 
symmetric model of \cite{Klebanov:2016xxf}, which appears to be
the minimal quantum mechanical model possessing the melonic limit.
\footnote{Our work may be generalized to similar models with higher rank tensors, but we won't do this explicitly here.} This model has $N^3$ anti-commuting degrees of freedom, $\psi^{abc}$, where
$a,b,c=1, \ldots, N$. In the model with global symmetry, the operators may be classified according to the group representations. In section \ref{composite} we study the spectra
of two-particle operators, which are either symmetric traceless or antisymmetric under two indices belonging to the same $O(N)$ group. We find that the spectrum
of symmetric traceless operators (\ref{nonsinglad}) is the same as that in the SYK model with real fermions; in particular it includes the $h=2$ zero-mode which plays an important role
in the dual gravitational dynamics \cite{Maldacena:2016upp,Engelsoy:2016xyb,Jensen:2016pah}. While in the SYK model there is one $h=2$ zero-mode, in
the $O(N)^3$ tensor model it appears with multiplicity $1+ \frac {3} {2} (N-1)(N+2)$. 
For the operators anti-symmetric in the two indices, (\ref{nonsinganti}), the spectrum is identical to the additional sector found in the complex tensor and
SYK models \cite{Klebanov:2016xxf,Fu:2016vas,Davison:2016ngz,Murugan:2017eto,Peng:2017spg,Bulycheva:2017uqj,Yoon:2017nig}; it includes the $h=0$ eigenvalue 
with multiplicity $\frac{3} {2} N (N-1)$ corresponding to
the conserved $O(N)^3$ charges.

An attractive feature of the tensor models is that the global symmetry may be gauged  \cite{Witten:2016iux,Klebanov:2016xxf}; this restricts the operator spectrum to
the invariant ones only. 
The ``Regge trajectory" of two-particle operators  $\psi^{abc}\partial_{t}^{2n+1}\psi^{abc}$ is clearly not the full set of $O(N)^3$ invariant
operators; there are vastly more operators which may be constructed by multiplying an even number of tensors and 
contracting all the indices \cite{Klebanov:2016xxf}. 
In section \ref{gaugeinv} we explicitly construct and draw pictorial representations of such operators 
(these pictures are analogous to the Feynman diagrams in the theory of three scalar fields $\varphi_i$ with interaction vertex $\varphi_1 \varphi_2 \varphi_3$).
Using the techniques developed in \cite{Sundborg:1999ue,Polyakov:2001af,Aharony:2003sx, Beccaria:2017aqc} (see also \cite{Boulatov:1991xz}),
we will calculate the number of $(2k)$-particle operators and 
show that it grows asymptotically as $2^k k!$. As a consequence, the theory
has a Hagedorn phase transition at the  temperature $\sim 1/\log N$, which we 
discuss in section \ref{sec:Hagedorn}.
Our work is similar in spirit to the classification of invariants in the $d=0$ tensor models \cite{Gurau:2011xp,Geloun:2013kta,Itoyama:2017xid,Mironov:2017aqv,Diaz:2017kub,deMelloKoch:2017bvv},
but some of our specific results appear to be new. Working with the quantum mechanical model of real 3-tensors introduces some subtleties and cancellations: for example, in the 
$O(N)^3$ fermionic model all the 6-particle operators vanish due to the Fermi statistics, while the number of $10$-particle operators is strongly reduced compared to the similar
bosonic model.  In section \ref{countinv} we also  count the invariants in $d=0$ bosonic models. In addition to the real tensors with $O(N)^3$ symmetry we study the complex tensor
theories with $U(N)^3$ and $U(N)^2 \times O(N)$ symmetries, as well as the symmetric traceless and fully antisymmetric rank-$3$ tensors under a single $O(N)$ group.

Beyond classifying the invariant operators, it is important to determine their infrared scaling dimensions. We begin work on this in section \ref{scalingdim} and point out that
there is a large class of $2k$-particle operators whose large $N$ scaling dimensions are simply additive, i.e. $k/2$. This is because the melonic ladders contribute only to $1/N$
corrections. However, although less generic, there are operators whose dimensions are not simply quantized. While the 
Regge trajectory operators studied in \cite{Kitaev:2015,Polchinski:2016xgd,Maldacena:2016hyu,Jevicki:2016bwu,Gross:2016kjj,Klebanov:2016xxf} receive single ladder
contributions, there are operators whose two-point functions have multi-ladder contributions. Since a ladder may contain an $h=2$ zero-mode, the $m$-ladder diagram seems to
produce a low-temperature enhancement by $(\beta J)^m$. This may
be an important
physical effect in the melonic tensor models, whose detailed analysis we leave for the future.

Besides our analysis of the spectra of $O(N)^3$ symmetric models, we make some comments about the $O(N)^6$ symmetric Gurau-Witten model  \cite{Witten:2016iux}.
Some features of its spectrum are identical to those in the $q=4$, $f=4$ Gross-Rosenhaus flavored generalization \cite{Gross:2016kjj} of the SYK model.
The connections of the Gurau-Witten model with this Gross-Rosenhaus model have been also noted using combinatorial analysis in \cite{Bonzom:2017pqs}.

After this paper was completed, we became aware of the interesting paper \cite{Choudhury:2017tax}, which has some overlap with our results. 

\section{Comments on the 
$O(N)^3$ Symmetric
Fermionic Tensor Quantum Mechanics }

Let us consider the quantum mechanical model of a real anticommuting 3-tensor $\psi^{abc}$ with the action \cite{Klebanov:2016xxf}
\begin{align}
S = \int d t \Big( \frac i 2 \psi^{abc}\partial_{t}\psi^{abc}+ \frac{1}{4}g \psi^{a_{1}b_{1}c_{1}}\psi^{a_{1}b_{2}c_{2}}\psi^{a_{2}b_{1}c_{2}}\psi^{a_{2}b_{2}c_{1}}\Big)\,. \label{FermAct1}
\end{align}
The three indices, each of which runs from $1$ to $N$, are treated as distinguishable, and the Majorana fermions satisfy the anti-commutation relations
\begin{align}
\{ \psi^{abc}, \psi^{a'b'c'}\} = \delta^{aa'} \delta^{bb'} \delta^{cc'}\ .
\label{anticom}
\end{align}
This model is a somewhat simplified version of the $O(N)^6$ symmetric Gurau-Witten model \cite{Witten:2016iux}. Both are in the class of 
3-tensor models which possess a ``melonic" large $N$ limit where $J= g N^{3/2}$ is held fixed 
\cite{Gurau:2009tw,Gurau:2011xp,Gurau:2011aq,Gurau:2011xq,Bonzom:2011zz,Tanasa:2011ur,Bonzom:2012hw,Dartois:2013he,Tanasa:2015uhr,Carrozza:2015adg,Gurau:2016lzk}.
The large $N$ model is nearly conformal in the IR \cite{Sachdev:1992fk,Kitaev:2015}; for example, the two-point function is
\begin{align}
\langle T(\psi^{abc}(t_1)\psi^{a'b'c'}(t_2))\rangle =  -\delta^{aa'}\delta^{bb'}\delta^{cc'} 
\Big(\frac{1}{4\pi g^{2}N^{3}}\Big)^{1/4}\ \frac{\sgn(t_{1}-t_{2})}{|t_{1}-t_{2}|^{1/2}}\,.
\label{IRtwopoint}
\end{align}

The model (\ref{FermAct1})
has the $O(N)_1\times O(N)_2 \times O(N)_3$ symmetry under the replacement\footnote{More generally, we could consider a model with $O(N_1)\times O(N_2) \times O(N_3)$ symmetry, where $a$ runs from $1$ to $N_1$, $b$ 
 from $1$ to $N_2$, and $c$ from $1$ to $N_3$. This may be thought of as a model of a large number $N_2$ of $N_1\times N_3$ matrices \cite{Ferrari:2017ryl}.}
\begin{align}
&\psi^{abc} \to M_{1}^{aa'} M_{2}^{bb'}M_{3}^{cc'}\psi^{a'b'c'}, \\
& M_{1}\in O(N)_1,\quad M_{2}\in O(N)_2, \quad M_{3} \in O(N)_3\,.
\end{align}
As far as the group $O(N)_1$ is concerned, we may think of $b$ and $c$ as flavor indices; therefore $\psi^{abc}$ produces $N^2$ flavors of real fermions in the fundamental of
$O(N)_1$. An analogous picture applies to $O(N)_2$ and $O(N)_3$.
The three sets of $SO(N)$ symmetry charges are
\begin{align}
Q_1^{a_{1} a_{2}}= \frac {i}{2} [\psi^{a_{1}b c }, \psi^{a_{2} b c}]\ , \quad 
Q_2^{b_{1} b_{2}}= \frac{i}{2} [\psi^{a b_1 c }, \psi^{a b_2 c}]\ , \quad 
Q_3^{c_{1} c_{2}}= \frac{i}{2} [\psi^{a b c_1 }, \psi^{a b c_2}]\,. \quad
\label{threecharges}
\end{align} 
The gauging of $SO(N)_1\times SO(N)_2 \times SO(N)_3$ sets these charges to zero; 
this restricts the operators to the invariant ones, 
where all the indices are contracted. In the ungauged model (\ref{FermAct1}) a more general class of operators is allowed, and they can be classified according to representations of the 
$SO(N)_1\times SO(N)_2 \times SO(N)_3$. 

Each $O(N)$ group
includes parity transformations (axis reflections) $P_{a_0}$: 
for a given $a_0$, $P_{a_0}$ sends 
$\psi^{a_0 b c} \rightarrow - \psi^{a_0 b c}$ for all $b,c$ and leaves all
$\psi^{a_1 b c}, a_1 \neq a_0$ invariant. In a physical
language, these are ``big'' gauge transformations and operators should 
be invariant under them. Therefore we can build operators 
using $\psi^{abc}$ and the delta symbol $\delta^{a a'}$ only. In 
the case of
$SO(N)$	gauge group one can use
the fully antisymmetric tensor $\epsilon_{a_1\dots a_N}$ as well; it is invariant under
$SO(N)$, but changes its sign under
the parity transformations. Because of this, there are additional ``long" operators containing at least $N$ fields, like
\begin{equation}
O_{\rm long}= \epsilon_{a_1 \dots a_N} \epsilon_{b_1 \dots b_N} \epsilon_{c_1 \dots c_N} \prod_{j=1}^N \psi^{a_j b_j c_j}\ .
\end{equation}
The difference
between gauging $O(N)$ and $SO(N)$ becomes negligible in the large $N$ limit.

Let us define three operations 
which permute pairs of the $O(N)$ symmetry groups (and thus interchange indices in the tensor field),
while also reversing the direction of time,
\begin{align}
&s_{ab}: \psi^{abc}\to \psi^{bac},  \qquad t\rightarrow -t ;\\
& s_{bc}:\psi^{abc} \to \psi^{acb}, \qquad t\rightarrow -t ;\\
& s_{ac}: \psi^{abc}\to \psi^{cba}, \qquad t\rightarrow -t \, .
\end{align}
Each of these transformations preserves the equations of motion
for the $\psi^{abc}$ field, 
\begin{align}
\dot{\psi}^{abc}= ig(\psi^{3})^{abc}\, , \quad (\psi^{3})^{abc}\equiv \psi^{a b_{1}c_{1}}\psi^{a_{1}b c_{1}}\psi^{a_{1}b_{1}c}\ .
\label{eom}
\end{align}
The Hamiltonian, including a quantum shift due to (\ref{anticom}), 
 \begin{align}
H=- \frac{1}{4}g \psi^{a_{1}b_{1}c_{1}}\psi^{a_{1}b_{2}c_{2}}\psi^{a_{2}b_{1}c_{2}}\psi^{a_{2}b_{2}c_{1}}+ \frac{gN^4}{16}=
- \frac{1}{4}g [\psi^{a_{1}b_{1}c_{1}}, \psi^{a_{1}b_{2}c_{2}}] [\psi^{a_{2}b_{1}c_{2}}, \psi^{a_{2}b_{2}c_{1}}]\ ,
\label{Hamilt}
\end{align}
changes sign under each of the transformations $s_{ab}$, $s_{bc}$, $s_{ac}$ (this is discussed in section \ref{gaugeinv}). 
This means that these transformations are unitary: they preserve $e^{i H t}$. In contrast, the usual time reversal transformation is anti-unitary because it also requires 
complex conjugation $i\rightarrow -i$. 

The $O(N)^3$ invariant operators form representations under the permutation group $S_3$, 
which acts on the three $O(N)$ symmetry groups (it contains the elements
$s_{ab}$, $s_{bc}$ and $s_{ac}$). For example, $H$ 
is in the degree $1$ "sign representation" of $S_3$: it changes sign under any pair interchange, but preserves its sign
under a cyclic permutation.

It is also interesting to study the spectrum of eigenstates of the Hamiltonian for small values of $N$; first steps on this were made in \cite{Klebanov:2017,Krishnan:2017ztz,Krishnan:2017txw}. 
When gauging the $O(N)^3$ symmetry one needs to worry about the $Z_2$ anomaly, which affects the gauged $O(N)$ quantum mechanics with an odd number of flavors of
real fermions in the fundamental
representation \cite{Witten:1985mj,Elitzur:1985xj}. Since for each of the three $O(N)$ groups we find $N^2$ flavors of fundamental fermions, 
the gauged model is consistent for even $N$, but is anomalous for odd $N$.\footnote{We are grateful to E. Witten for pointing this out to us.}
This means that, for odd $N$, the spectrum does not contain states which are invariant under $O(N)^3$ (for $N=3$ this can be seen via an explicit diagonalization of the Hamiltonian
 (\ref{Hamilt}) \cite{Klebanov:2017}).

\section{Composite Operators and Schwinger-Dyson Equations}
\label{composite}

The scaling dimensions of a class of bilinear operators may be extracted from the 4-point function \cite{Klebanov:2016xxf}
\begin{align}
\langle \psi^{a_1 b_1 c_1 } (t_1)  \psi^{a_1 b_1 c_1 } (t_2) \psi^{a_2 b_2 c_2 } (t_3)  \psi^{a_2 b_2 c_2 } (t_4) \rangle \,,
\end{align}
and factorizing it in the channel where $t_1\rightarrow t_2$ and $t_3\rightarrow t_4$. A class of melonic ladder graphs appears in this channel in the large $N$
limit; it may be summed
by means of a Schwinger-Dyson equation.
The singlet bilinear operators 
\begin{equation}
{\mathcal O}_n=
\psi^{abc}\partial_{t}^{2n+1}\psi^{abc}\ , \qquad n=0,1,2, \ldots
\label{singletbilin}
\end{equation}
 form a ``Regge trajectory." Their scaling dimensions are the same as in the SYK model \cite{Sachdev:1992fk,Kitaev:2015}, and they
have been extensively analyzed in the literature \cite{Polchinski:2016xgd,Maldacena:2016hyu,Jevicki:2016bwu,Gross:2016kjj}.
The dimensions are determined by the equation
\begin{align}
g (h)=  -\frac{3}{2}\frac{\tan(\frac{\pi}{2}(h-\frac{1}{2}))}{h-1/2}=1\ ,
\label{gantisym}
\end{align}
and the first few solutions are $h=2, 3.77, 5.68, \ldots$. 
As pointed out in \cite{Klebanov:2016xxf}, the model also contains a multitude of multi-particle singlet operators. As we will see, some special combinations of the multi-particle operators
are related by the equations of motion to the operators (\ref{singletbilin}), but most multi-particle operators are genuinely new.

\begin{figure}[h!]
                \centering
                \includegraphics[width=6cm]{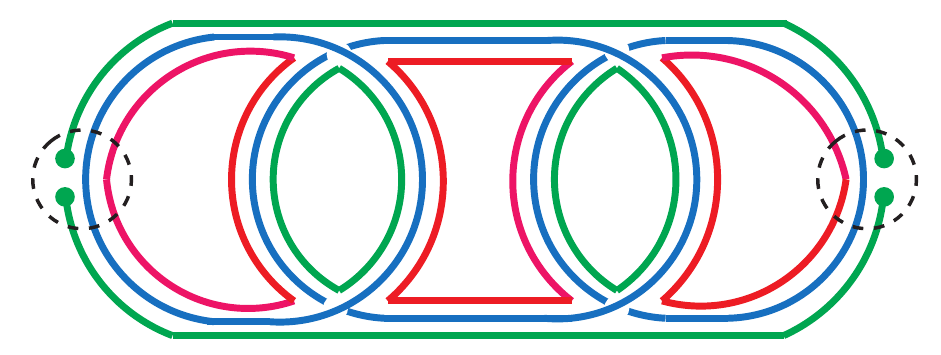}
                \caption{A ladder contribution to the two-point function of a bilinear operator with two pairs of indices contracted, $\mathcal O^{c_1 c_2}$. It is not suppressed in the large $N$ limit. }
                \label{QQ}
\end{figure}

Interestingly, there are also certain non-singlet operators which are renormalized by the melonic ladder diagrams.
This can be seen, for example, from the 4-point function
\begin{align}
\langle \psi^{a_1 b_1 c_1 } (t_1)  \psi^{a_2 b_1 c_1 } (t_2) \psi^{a_1 b_2 c_2 } (t_3)  \psi^{a_2 b_2 c_2 } (t_4) \rangle 
\label{nonsing}
\end{align}
factorized in the channel $t_1\rightarrow t_2$ and $t_3\rightarrow t_4$. As shown in figure \ref{QQ}, all the melonic ladders again make non-vanishing contributions in the
large $N$ limit. Here we find two classes of non-singlet bilinear operators: those symmetric and traceless in $a_1$ and $a_2$, and those anti-symmetric.
The $\frac {1} {2} (N-1)(N+2)$ symmetric traceless operators under $O(N)_1$,
\begin{align}
{\mathcal O}_n^{(a_1 a_2)}= \psi^{a_1bc}\partial_{t}^{2n+1}\psi^{a_2bc}+ \psi^{a_2 bc}\partial_{t}^{2n+1}\psi^{a_1 bc}- \frac 2 N \delta^{a_1 a_2}  \psi^{abc}\partial_{t}^{2n+1}\psi^{abc}
\ , 
\label{nonsinglad}
\end{align}
where $n=0,1,2, \ldots$, have the same spectrum as the singlet bilinears (\ref{singletbilin}) which is determined by
(\ref{gantisym}).
Of course, there are analogous operators 
${\mathcal O}_n^{(b_1 b_2)}$ and ${\mathcal O}_n^{(c_1 c_2)}$
that are symmetric traceless under $O(N)_2$ and $O(N)_3$, respectively.
Thus, the symmetric traceless operators present in the ungauged model contain the $h=2$ zero-mode with multiplicity  $\frac {3} {2} (N-1)(N+2)$; this 
appears to imply a significant physical difference between the ungauged
$O(N)^3$ model and the SYK model.\footnote{We are grateful to Shiraz Minwalla for very useful discussions on this; see the paper \cite{Choudhury:2017tax}.}
While in the gauged model such bilinear operators are projected out, we may form singlet combinations out of their products; such operators have an interesting feature
that they are renormalized by multiple ladders. For example, in section \ref{scalingdim} we will encounter operators related by the equation of motion to
${\mathcal O}_0^{(a_1 a_2)} {\mathcal O}_0^{(a_1 a_2)}$, so they are renormalized by double ladders. The pictorial representations
of these operators may be found in column 2 of figure \ref{O8all}.

There are also the $\frac {1} {2} N(N-1)$ operators in the anti-symmetric two-index representation of $O(N)_1$,
\begin{align}
{\mathcal O}_n^{[a_1 a_2]}=
\psi^{a_1bc}\partial_{t}^{2n}\psi^{a_2bc}- \psi^{a_2 bc}\partial_{t}^{2n}\psi^{a_1 bc}\ ,
\label{nonsinganti}
\end{align} 
and the analogous anti-symmetric operators under $O(N)_2$ and $O(N)_3$. The Schwinger-Dyson equations for these operators are identical to the "symmetric sector" of
the complex tensor model \cite{Klebanov:2016xxf,Fu:2016vas,Davison:2016ngz,Murugan:2017eto,Peng:2017spg,Bulycheva:2017uqj,Yoon:2017nig}. 
Their scaling dimensions are determined by
\begin{align}
\tilde g (h)= -\frac{1}{2}\frac{\tan(\frac{\pi}{2}(h+\frac{1}{2}))}{h-1/2}=1\, .
\label{gsym}
\end{align}
The first few solutions of this equation are $h=0, 2.65, 4.58, \ldots$, and each one appears with multiplicity $\frac {3} {2} N(N-1)$. 
The spectrum includes the special $h=0$ mode corresponding here to the $n=0$ operators, which are the $O(N)^3$ charges (\ref{threecharges}).

\begin{figure}[h!]
                \centering
                \includegraphics[width=12cm]{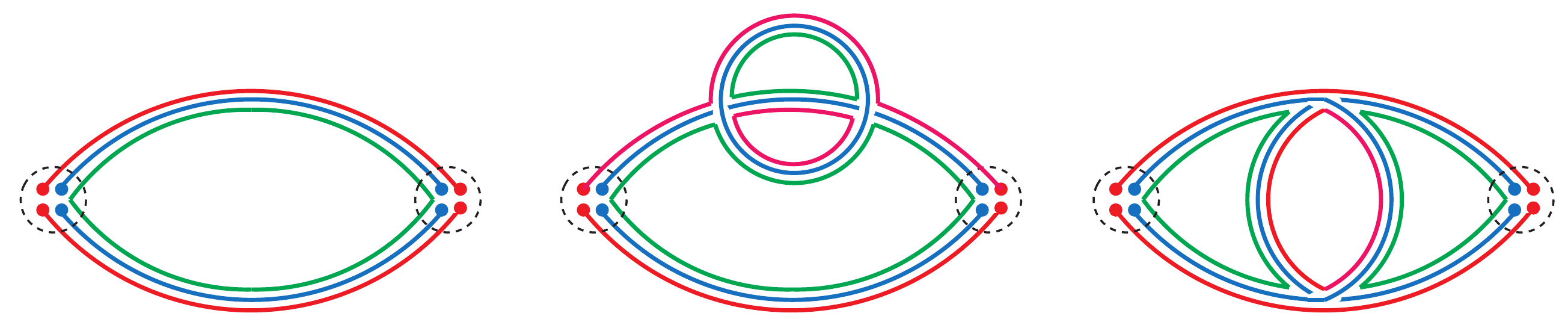}
                \caption{Different contributions to the two-point function of a bilinear operator with one pair of indices contracted, ${\mathcal O}_m^{b_1 c_1 b_2 c_2}$. The ladder diagrams, such as the rightmost figure, are suppressed in the large $N$ limit. }
                \label{QQ1}
\end{figure}

The 4-point function (\ref{nonsing}) may also be factorized in the channel $t_1\rightarrow t_3$ and $t_2\rightarrow t_4$. This leads to the spectrum of operators
\begin{align}
{\mathcal O}_m^{b_1 c_1 b_2 c_2}= \psi^{a b_1 c_1 }\partial_{t}^{m}\psi^{a b_2 c_2}\ .
\label{fourindex}
\end{align}
We can see from figure \ref{QQ1} that the ladder contribution to this operator are subleading in $1/N$: the rightmost diagram is of ladder type and 
 is $\sim g^2 N^3$, which is suppressed by a power of $N$ relative to the other two diagrams. 
Therefore the large $N$ scaling dimensions of these operators are $1/2 + m$.

\begin{figure}[h!]
                \centering
                \includegraphics[width=11cm]{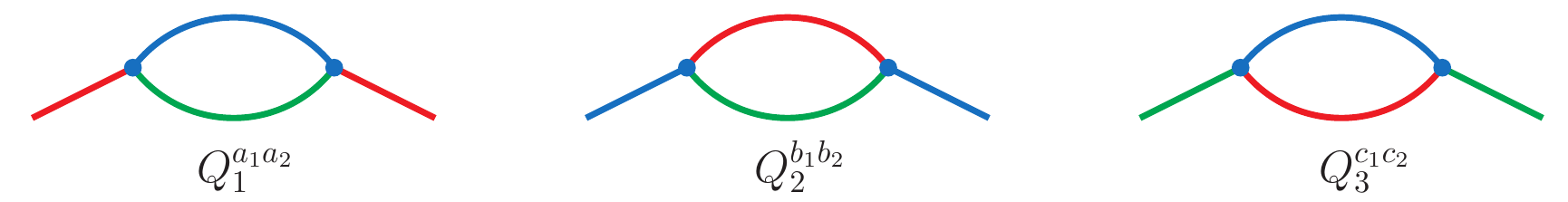}
                \caption{The $O(N)_1$, $O(N)_2$ and $O(N)_3$ charges.}
                \label{Charges}
\end{figure}

We will adopt a pictorial representation of the operators where the $\psi^{abc}$ fields are shown as the vertices. 
The $a$-indices which transform under $O(N)_1$ are shown by red lines; 
the $b$-indices which transform under $O(N)_2$ are shown by blue lines; and the $c$-indices which transform under $O(N)_3$ are shown by green lines.
For example, the three charges (\ref{threecharges}) are shown in figure \ref{Charges}.

\section{Construction of $O(N)^3$ invariant operators}
\label{gaugeinv}

\begin{figure}[h!]
                \centering
                \includegraphics[width=13cm]{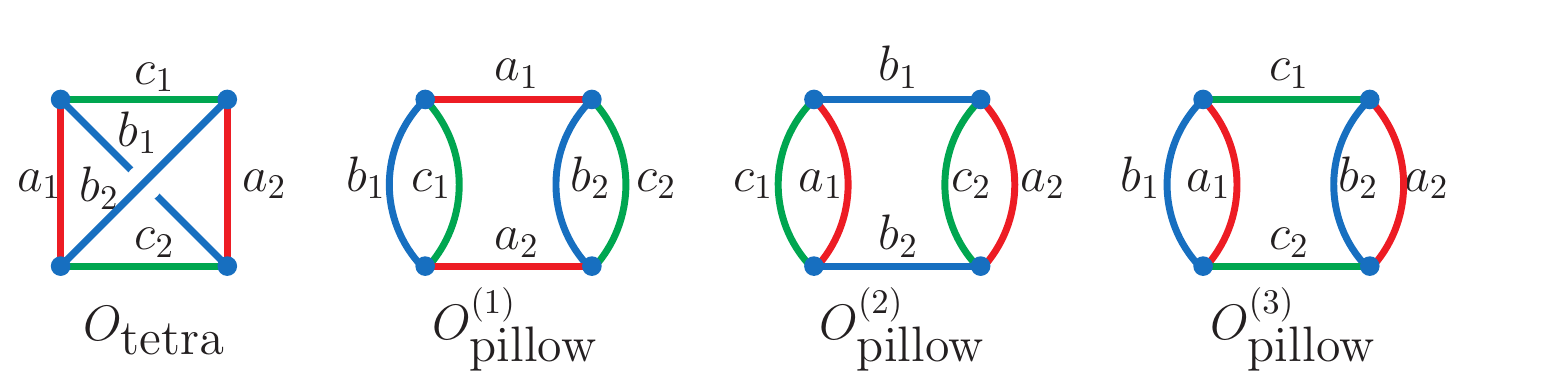}
                \caption{All the four-particle operators, the tetrahedron and the three pillows, with the index contractions shown explicitly.}
                \label{O4ops}
\end{figure}

In this section we study the spectrum of $O(N)^3$ invariant operators. 
Since a time derivative may be removed using the equations of motion (\ref{eom}), we may write the operators in a form where no derivatives are present.
The bilinear singlet operator, $\psi^{abc}\psi^{abc}$, vanishes classically by the Fermi statistics, 
while at the quantum level taking into account (\ref{anticom}), it is a C-number. The first non-trivial operators appear at the quartic level and are shown in figure \ref{O4ops} 
(from here on we will not be careful about the quantum corrections to operators).

On the left is the 
``tetrahedron operator" $O_{\textrm{tetra}}$, which is proportional to the Hamiltonian (\ref{Hamilt}):
\begin{align}
O_{\textrm{tetra}} = \psi^{a_{1}b_{1}c_{1}}\psi^{a_{1}b_{2}c_{2}}\psi^{a_{2}b_{1}c_{2}}\psi^{a_{2}b_{2}c_{1}}\,.
\end{align}
One can check that 
\begin{align}
s_{bc}O_{\textrm{tetra}} &
 =\psi^{a_{1}c_{1}b_{1}}\psi^{a_{1}c_{2}b_{2}}\psi^{a_{2}c_{2}b_{1}}\psi^{a_{2}c_{1}b_{2}} \notag\\
&=\psi^{a_{1}b_{1}c_{1}}\psi^{a_{1}b_{2}c_{2}}\psi^{a_{2}b_{2}c_{1}}\psi^{a_{2}b_{1}c_{2}} = -O_{\textrm{tetra}} \,,
\end{align}
and also that  $s_{ab}O_{\textrm{tetra}} = -O_{\textrm{tetra}} $ and $s_{ac}O_{\textrm{tetra}} = -O_{\textrm{tetra}} $.
Thus, the tetrahedron operator $O_{\textrm{tetra}}$ is in the degree $1$ "sign representation" of $S_3$: it changes sign under any pair interchange, but preserves its sign
under a cyclic permutation.

The three additional operators in figure \ref{O4ops}, which we denote as $O_{\textrm{pillow}}^{(1)}$,  $O_{\textrm{pillow}}^{(2)}$ and $O_{\textrm{pillow}}^{(3)}$, are the "pillow" operators in the terminology of
\cite{Tanasa:2011ur,Carrozza:2015adg}; they
contain double lines between a pair of vertices.
For example, for  $O_{\textrm{pillow}}^{(1)}$ we have
\begin{align}
O_{\textrm{pillow}}^{(1)} =-\psi^{a_{1}b_{1}c_{1}}\psi^{a_{2}b_{1}c_{1}}\psi^{a_{1}b_{2}c_{2}}\psi^{a_{2}b_{2}c_{2}}=Q_1^{a_{1} a_{2}} Q_1^{a_{1} a_{2}} \,.
\end{align}
Under the $S_3$ the three pillow operators decompose into the trivial representation of degree $1$ and the standard representation of degree $2$.
Since the charges (\ref{threecharges}) commute with the Hamiltonian (\ref{Hamilt}), so does each of the three pillow operators. 
This means that the scaling dimensions of the pillow operators are unaffected by the interactions, i.e. they vanish. In fact, the three pillow operators are simply
the quadratic Casimir operators of the three $O(N)$ groups.\footnote{We thank Dan Roberts and Douglas Stanford for discussions on this.} 
The gauging of $O(N)^3$ symmetry sets the charges (\ref{threecharges}) to zero, so the pillow operators do not appear in the gauged model.

Using the equations of motion (\ref{eom}) we see that the operator 
$O_{\textrm{tetra}}$ is related by the equation of motion to the operator $\psi^{abc}\partial_{t}\psi^{abc}$
\begin{align}
O_{\textrm{tetra}} = \psi^{abc} (\psi^{3})^{abc}  \propto \psi^{abc}\partial_{t}\psi^{abc}\,.
\end{align}
If we iterate the use of the equation of motion (\ref{eom}), then all derivatives in an operator may be traded for extra $\psi$-fields. 
Thus, a complete basis of operators may be constructed by
multiplying some number $2k$ of $\psi$-fields and contracting all indices. In this approach, there is a unique operator with $k=2(n+1)$ which is equal to the
Regge trajectory operator $\psi^{abc}\partial_{t}^{2n+1}\psi^{abc}$. For $n=0$ this operator is $O_{\textrm{tetra}}$, which is proportional to the Hamiltonian; for $n=1$ it will be constructed
explicitly in section \ref{eightpart}.

 \begin{figure}[h!]
                \centering
                \includegraphics[width=10cm]{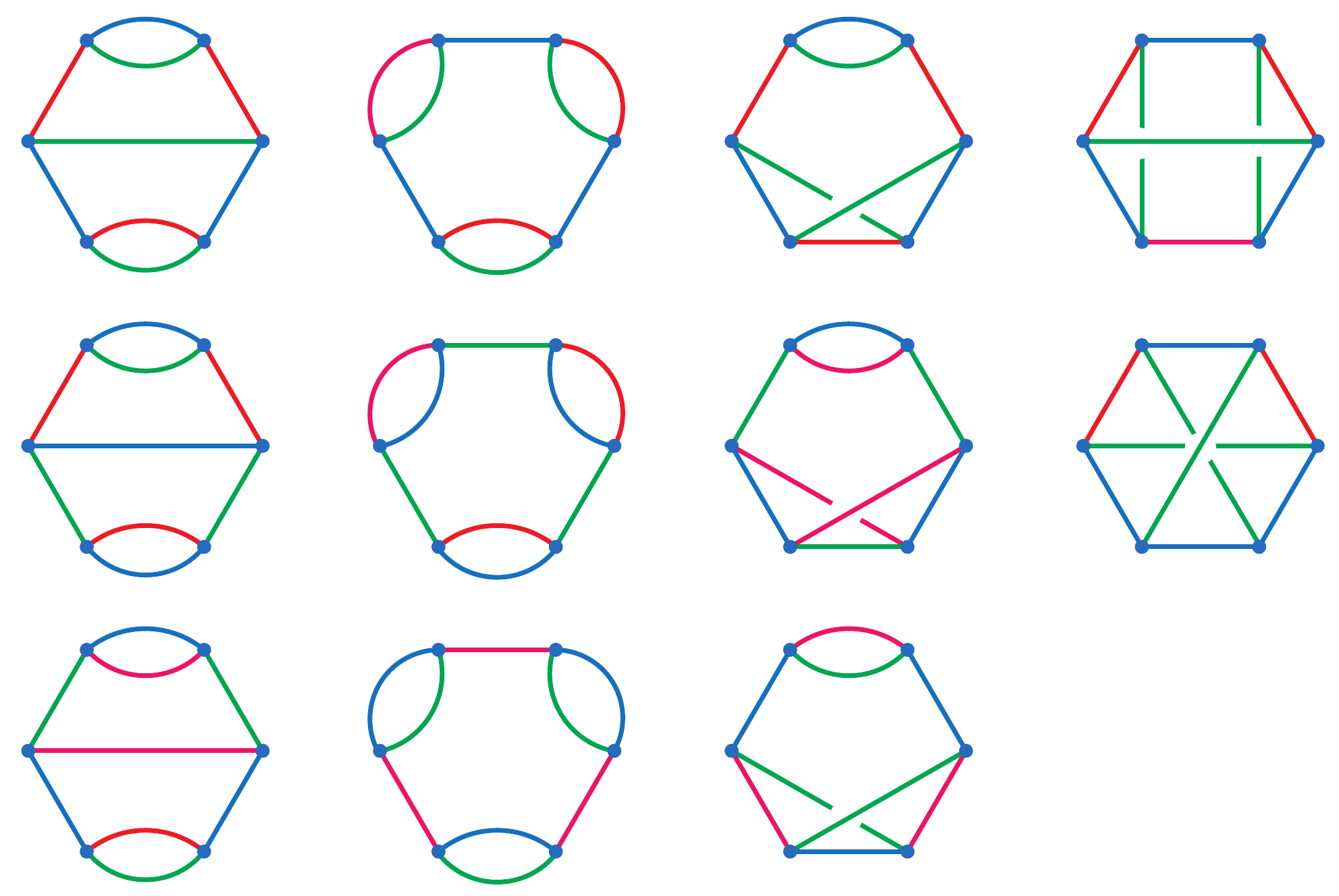}
                \caption{All  six-particle operators. They are present in the scalar model but vanish in the fermionic model.}
                \label{O6ops}
\end{figure} 

All the six-particle operators are represented in figure \ref{O6ops}, but
due to the Fermi statistics all of them vanish. Even if this were not the case, the operators in the first three columns would vanish in the gauged model 
because they contain insertions of the charges (\ref{threecharges}). Let us demonstrate the vanishing of the  two operators in the last column in detail.  
The first operator
\begin{align}
O_{6}^{(1)} =\psi^{a_{1}b_{1} c_{1}}\psi^{a_{1}b_{2}c_{2}}\psi^{a_{2} b_{1}c_{2}}\psi^{a_{2}b_{3}c_{3}}\psi^{a_{3}b_{3}
  c_{1}}\psi^{a_{3}b_{2}c_{3}} \,,
\end{align}
may be written as 
\begin{align}
O_{6}^{(1)} =(\psi^{3})^{a_{2}b_{2}c_{1}}(\psi^{3})^{a_{2}b_{2}c_{1}} = 0\,.
\end{align}
This may be seen by cutting the diagram for this operator in figure \ref{O6ops} along the vertical symmetry axis.
To show that 
\begin{align}
O_{6}^{(2)} =\psi^{a_{1}b_{1}c_{1}}\psi^{a_{1}b_{2}c_{2}}\psi^{a_{2}b_{2}c_{3}}\psi^{a_{2}b_{3}c_{1}}\psi^{a_{3}b_{3} 
  c_{2}}\psi^{a_{3}b_{1}c_{3}} \, 
\end{align}
also vanishes, we may permute the first two $\psi$-fields to write it as
\begin{align}
O_{6}^{(2)} =-\psi^{a_{1}b_{2}c_{2}}\psi^{a_{1}b_{1}c_{1}}\psi^{a_{2}b_{2}c_{3}}\psi^{a_{2}b_{3}c_{1}}\psi^{a_{3}b_{3} 
  c_{2}}\psi^{a_{3}b_{1}c_{3}} \ .
\end{align}
After relabeling $b_1\leftrightarrow b_2$, $c_1\leftrightarrow c_2$ and $a_2\leftrightarrow a_3$, we observe that the RHS equals $-O_{6}^{(2)}$.
Therefore, $O_{6}^{(2)}=- O_{6}^{(2)}=0$.

\begin{figure}[h!]
                \centering
                \includegraphics[width=13cm]{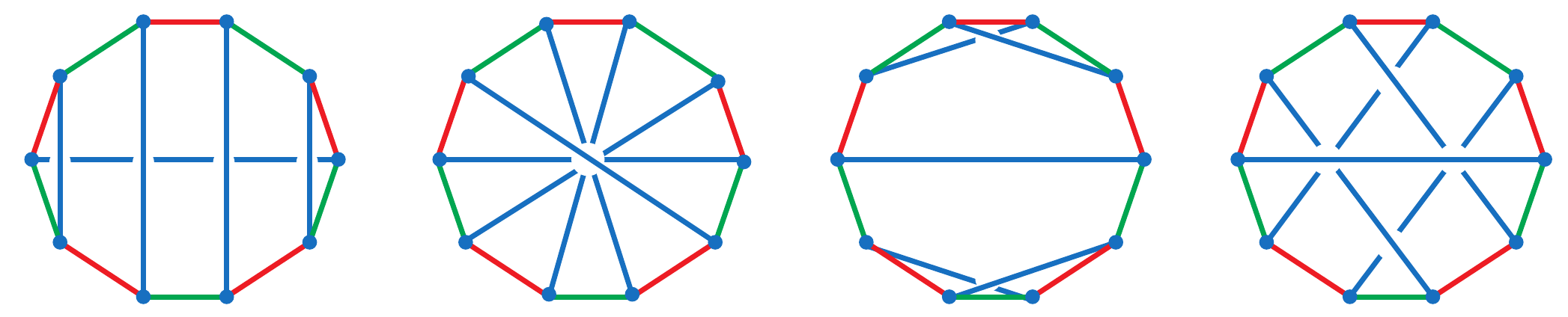}
                \caption{Some ten-particle operators which vanish in the fermionic model.}
                \label{O10}
\end{figure} 

One may wonder if the vanishing extends to the $10$-particle operators. We have checked that the operators shown in 
figure \ref{O10} all vanish; this is due to the reflection symmetry present for these operators. 
For example, the left operator in figure \ref{O10} vanishes because it may be written as
$(\psi^5)^{abc} (\psi^5)^{abc}$, 
which may be seen by cutting the diagram along the vertical symmetry axis.
We note that
\begin{align}
(\psi^5)^{abc}= g^{-2} \partial_t^2 \psi^{abc}\, .
\label{eom2}
\end{align}
Similarly, by cutting the third diagram in figure \ref{O10} along its vertical symmetry axis, we see that the corresponding operator
may be written as $(\psi^5)^{a b_1 b_2 b_3 b_4 c} (\psi^5)^{a b_1 b_2 b_3 b_4 c}$ which obviously vanishes as well. 
This argument extends to all the reflection symmetric $(4n+2)$-particle diagrams.

\begin{figure}[h!]
                \centering
                \includegraphics[width=6.5cm]{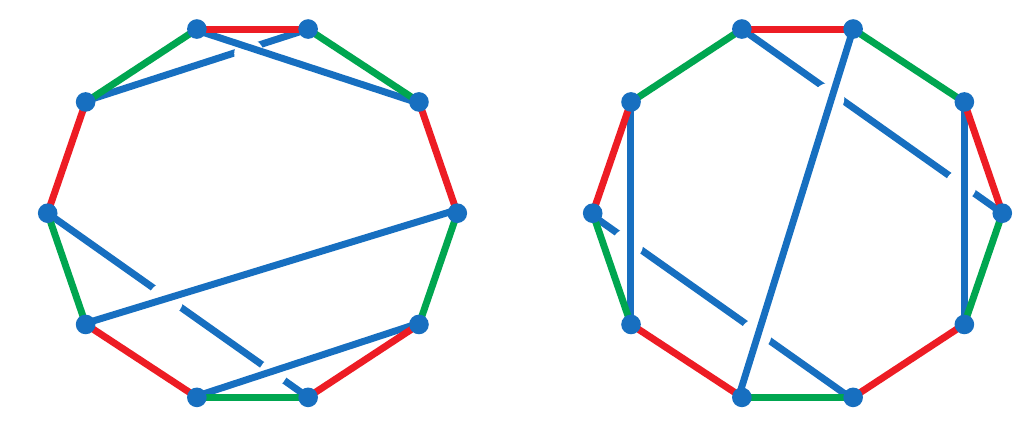}
                \caption{Some non-vanishing ten-particle operators.}
                \label{O10nonzero}
\end{figure}

However, not all $10$-particle operators vanish. For example, the operators shown in figure \ref{O10nonzero} do not have a reflection symmetry, 
and we have checked that they do not vanish.

Let us note that each gauge invariant operator, where all the indices are contracted, corresponds to a vacuum Feynman diagram in the 
theory with three scalar fields and interaction $\lambda \varphi_1 \varphi_2 \varphi_3$ (the three different propagators correspond to the lines of three different colors
in our figures).
In the theory of bosonic tensors $\phi^{abc}$, the number of operators made out of $2k$ fields is precisely the number of distinct Feynman 
diagrams appearing at order $\lambda^{2k}$, which grows as $k! 2^k$. In the fermionic model, some of the operators vanish by the Fermi statistics,
while others due to the gauge constraint. Nevertheless, we will find that the factorial growth holds also in the fermionic model.

\subsection{Eight-particle operators}
\label{eightpart}

\begin{figure}[h!]
                \centering
                \includegraphics[width=12cm]{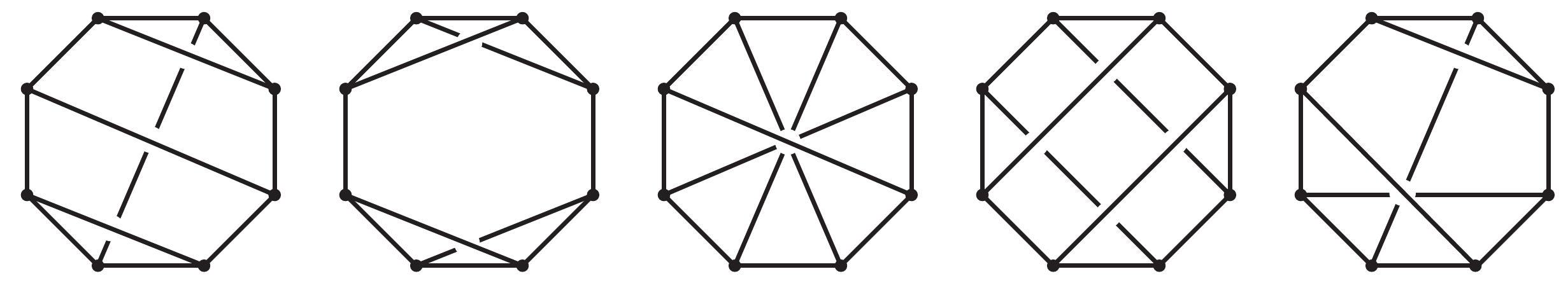}
                \caption{Eight-particle operator topologies}
                \label{O8n1}
\end{figure}

\begin{figure}[h!]
                \centering
                \includegraphics[width=15cm]{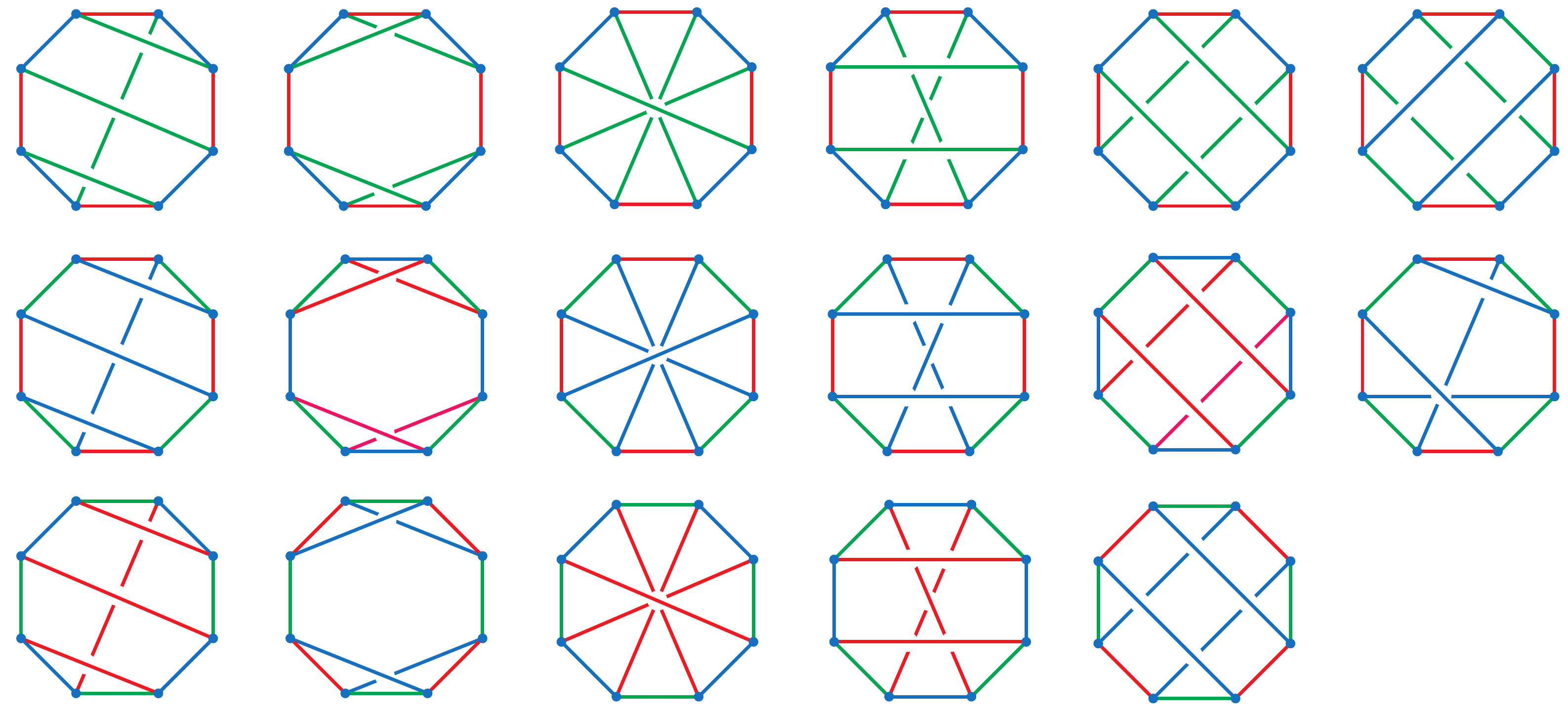}
                \caption{All eight-particle operators in the fermionic model.}
                \label{O8all}
\end{figure}

In this section we explicitly construct all the eight-particle operators without bubble (double line) insertions and exhibit their pictorial representations. 
Having two vertices connected by a double line corresponds to insertion of an
$O(N)$ charge which vanishes in the gauged model. For this reason we will omit such operators and list only those where there are no double lines.
The possible topologically inequivalent eight-particle operators are shown in figure \ref{O8n1}; from these we can obtain other admissible operators by
interchanging the colors. In this way we find 17 inequivalent operators shown in figure \ref{O8all}.

Among the eight-particle operators there are three which may be obtained from the tetrahedral vertex 
\begin{align}
 &O_{1}= \psi^{a_{1}b_{1}c_{1}}\psi^{a_{1}b_{2}c_{2}}\psi^{a_{2}b_{2}c_{1}}\psi^{a_{2}b_{4}c_{4}}\psi^{a_{3}b_{3}
  c_{2}}\psi^{a_{3}b_{1}c_{3}}\psi^{a_{4}b_{4}c_{3}}\psi^{a_{4}b_{3} c_{4}}\,,\notag\\
 &O_{2}=  \psi^{a_{1}b_{1} c_{1}}\psi^{a_{1} b_{2} c_{2}}\psi^{a_{2}b_{2} c_{1}}\psi^{a_{2}b_{3} c_{3}}\psi^{a_{3}b_{3}
  c_{2}}\psi^{a_{3} b_{4}c_{4}}\psi^{a_{4} b_{4}c_{3}}\psi^{a_{4} b_{1}c_{4}} \,, \label{tetrahedral}\\
  &O_{3}=\psi^{a_{1}b_{1} c_{1}}\psi^{a_{1}b_{2}c_{2}}\psi^{a_{2} b_{2} c_{1}}\psi^{a_{2} b_{3} c_{3}}\psi^{a_{3} b_{1}
  c_{3}}\psi^{a_{3} b_{4} c_{4}}\psi^{a_{4} b_{3} c_{4}}\psi^{a_{4} b_{4}c_{2}} \,.\notag
\end{align}
Their pictorial representations are shown in the first column of figure \ref{O8all}. 
Using the equations of motion, we may write them as
\begin{align}
& O_1= \dot \psi^{a_{1}b_{1}c_{1}}\dot \psi^{a_{1}b_{2}c_{2}}\psi^{a_{2}b_{1}c_{2}} \psi^{a_{2}b_{2}c_{1}}\,,\notag\\
& O_2= \dot \psi^{a_{1}b_{1}c_{1}}\psi^{a_{1}b_{2}c_{2}}\dot \psi^{a_{2}b_{1}c_{2}} \psi^{a_{2}b_{2}c_{1}}\,,\notag\\
& O_3= \dot \psi^{a_{1}b_{1}c_{1}}\psi^{a_{1}b_{2}c_{2}} \psi^{a_{2}b_{1}c_{2}} \dot \psi^{a_{2}b_{2}c_{1}}\, .
\end{align}
It follows that
\begin{align}
O_{1}+O_{2}+O_{3}\sim \partial_t \psi^{abc} \partial_{t}^{2}\psi^{abc}
\ ,
 \label{melonic_O8}
\end{align}
which up to a total derivative equals the Regge trajectory operator $ \psi^{abc}\partial_{t}^{3}\psi^{abc}$.

The transformation properties of operators $O_{1}$, $O_{2}$ and $O_{3}$ under $S_3$ are
\begin{align}
& s_{bc}O_{3}=-O_{2}, \quad s_{bc}O_{2}=-O_{3},\quad s_{bc}O_{1}=-O_{1}\ ,\notag \\
&s_{ac}O_{3}=-O_{1}, \quad s_{ac}O_{2}=-O_{2},\quad s_{ac}O_{1}=-O_{3}\,,\notag\\
&s_{ab}O_{3}=-O_{3}, \quad s_{ab}O_{2}=-O_{1},\quad s_{ab}O_{1}=-O_{2}\, .\notag
\end{align}
It follows that 
\begin{align}
(s_{ab},s_{ac}, s_{bc}): (O_{1}+O_{2}+O_{3})\to -(O_{1}+O_{2}+O_{3})\,.
\end{align}
Therefore, the operator $\psi^{abc}\partial_{t}^{3}\psi^{abc}\sim O_{1}+O_{2}+O_{3}$ is in the degree 1 sign representation of $S_3$.
The other two linear combinations of operators (\ref{tetrahedral}), $O_1- O_2$ and $O_2-O_3$, form the standard degree 2 representation of $S_3$. 

Similarly, we may write down the three operators which correspond to the 
second column in figure \ref{O8all} (the first of these operators, $\tilde O_{1}$, was written down in \cite{Klebanov:2016xxf}):
\begin{align}
 &\tilde O_{1}= \psi^{a_{1} b_{1}c_{1}}\psi^{a_{1} b_{2} c_{2}}\psi^{a_{2} b_{3} c_{3}}\psi^{a_{2} b_{4} c_{4}}\psi^{a_{3} b_{1} 
  c_{3}}\psi^{a_{3} b_{3} c_{1}}\psi^{a_{4}b_{2} c_{4}}\psi^{a_{4} b_{4} c_{2}}\,,\notag\\
 &\tilde O_{2}=  \psi^{a_{1}b_{1}c_{1}}\psi^{a_{2} b_{1} c_{2}}\psi^{a_{3} b_{2} c_{3}}\psi^{a_{4}b_{2} c_{4}}\psi^{a_{1} b_{3} 
  c_{3}}\psi^{a_{3} b_{3} c_{1}}\psi^{a_{2} b_{4} c_{4}}\psi^{a_{4} b_{4} c_{2}} \,,\label{ourops}\\
  &\tilde O_{3}=\psi^{a_{1} b_{1} c_{1}}\psi^{a_{2} b_{2} c_{1}}\psi^{a_{3} b_{3} c_{2}}\psi^{a_{4} b_{4} c_{2}}\psi^{a_{3} b_{1} 
  c_{3}}\psi^{a_{1} b_{3} c_{3}}\psi^{a_{4} b_{2} c_{4}}\psi^{a_{2} b_{4} c_{4}} \,. \notag
\end{align}
Via the equations of motion, these operators are related to the bilinear operators defined in (\ref{nonsinglad}):
\begin{equation}
\tilde O_{1} \sim {\cal O}_0^{(a_1 a_2)} {\cal O}_0^{(a_1 a_2)}\ ,\qquad
\tilde O_{2} \sim {\cal O}_0^{(b_1 b_2)} {\cal O}_0^{(b_1 b_2)}\ ,\qquad
\tilde O_{3} \sim {\cal O}_0^{(c_1 c_2)} {\cal O}_0^{(c_1 c_2)}\ .
\end{equation}
These relations will be used in the next section.

The action of the discrete symmetries on the operators is
\begin{align}
&s_{bc} \tilde O_{3}=\tilde O_{2}, \quad s_{bc} \tilde O_{2}=\tilde O_{3},\quad s_{bc} \tilde O_{1}=\tilde O_{1}\,,\notag\\
&s_{ac}\tilde O_{3}=\tilde O_{1}, \quad s_{ac}\tilde O_{2}=\tilde O_{2},\quad s_{ac}\tilde O_{1}=\tilde O_{3}\,,\notag\\
&s_{ab}\tilde O_{3}=\tilde O_{3}, \quad s_{ab}\tilde O_{2}=\tilde O_{1},\quad s_{ab}\tilde O_{1}=\tilde O_{2}\,,
\end{align}
so that 
\begin{align}
(s_{ab},s_{bc}, s_{ac}): \tilde O_{1}+\tilde O_{2}+ \tilde O_{3}\to \tilde O_{1}+ \tilde O_{2}+ \tilde O_{3}\,.
\end{align}
Therefore, this operator is in the trivial representation of $S_3$.
The other two linear combinations of operators (\ref{ourops}), $\tilde O_1- \tilde O_2$ and $\tilde O_2-\tilde O_3$, form the standard degree $2$ representation of $S_3$. 
The operators corresponding to the other topologies in figure \ref{O8n1} may be written down analogously.

\section{Scaling Dimensions of Multi-Particle Operators}
\label{scalingdim}

\begin{figure}[h!]%
    \centering
  \includegraphics[width=12cm]{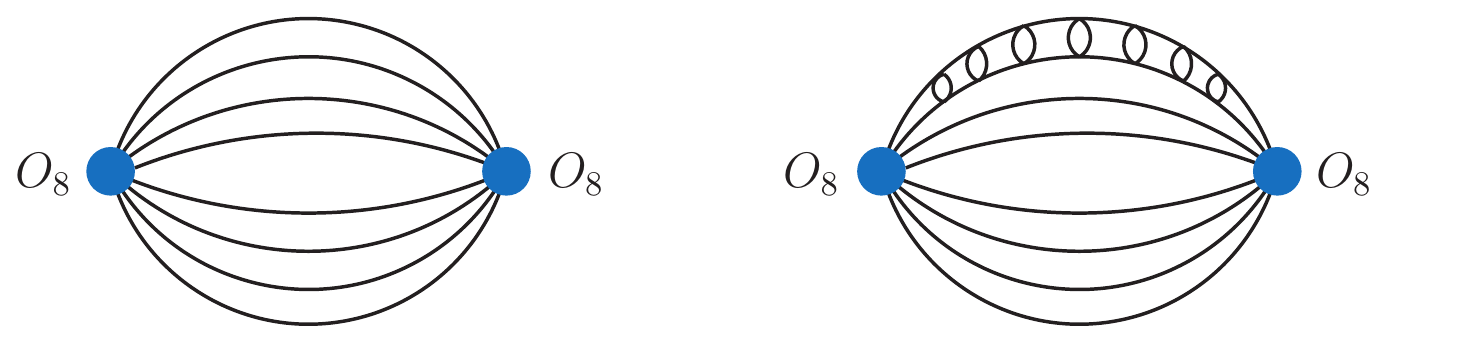}%
    \caption{Diagrammatics for the ``typical' operators whose IR dimensions are quantized. Each line denotes a dressed propagator. 
a) The melonic diagrams that contribute to the operator two-point functions in the large $N$ limit. b) The ladder diagrams which do not contribute in the large $N$ limit.}
    \label{LOrder}%
\end{figure}

We have seen that the tensor models admit a variety of singlet operators. In this section we discuss their scaling dimensions.
Since operators ${\mathcal O}_m^{b_1 c_1 b_2 c_2}$ defined in (\ref{fourindex}) do not receive ladder contributions in the large $N$ limit,
we expect a large class of $m$-particle operators to have the quantized dimensions:\footnote{We are very grateful to E.~Witten for pointing this out to us.}

\begin{equation}
    \Delta_m=\frac{m}{4} +\mathcal O\left( 1/N \right).
    \label{Delta_m}
\end{equation}
This is the dimension of an operator which is not renormalized by ladder diagrams because every pair of tensors have at most one index in common.
This situation is illustrated in figure \ref{LOrder}: the dominant contribution comes from the two operators contracted using the IR two-point function
(\ref{IRtwopoint}), and the ladder insertions are suppressed by $1/N$.
We find that this applies to most of the 17 eight-particle operators shown in figure \ref{O8all}. The exceptions are operators 
$O_i$ and $\tilde O_{i}$, defined in (\ref{tetrahedral}), (\ref{ourops}), and shown
in columns 1 and 2.
For example, each of the operators $\tilde O_{i}$ in column 2 is renormalized by two ladders, as we discuss below.

Thus, the $m/4$ rule does not apply to all operators: it is violated for the operators whose two-point functions receive the melonic ladder contributions in the large $N$
limit.
One class of such singlet operators is the Regge trajectory we have discussed before:
\begin{equation}
    \psi^{abc} \partial_t^{2n+1} \psi^{abc}.
    \label{regge_def}
\end{equation}
After applying the equation of motion (\ref{eom}), which
schematically may be represented as
\begin{equation} 
    \includegraphics[width=8cm]{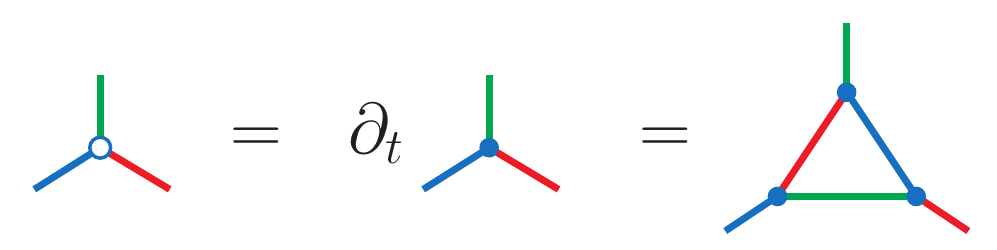}
    \label{eom_schem}
\end{equation}
we may represent the Regge trajectory operators in terms of multi-particle operators without derivatives. For example, the $n=0$ operator is equivalent to the
4-particle ``tetrahedron" operator $O_{\textrm{tetra}}$, while the  $n=1$ operator is equivalent to $O_1+ O_2+ O_3$, as shown in
(\ref{melonic_O8}). 
The dimensions of such operators come from solving (\ref{gantisym}), so the operator $O_1+ O_2+ O_3$ has $h\approx 3.77$.

\begin{figure}[h!]
\centering
    \includegraphics[width=14cm]{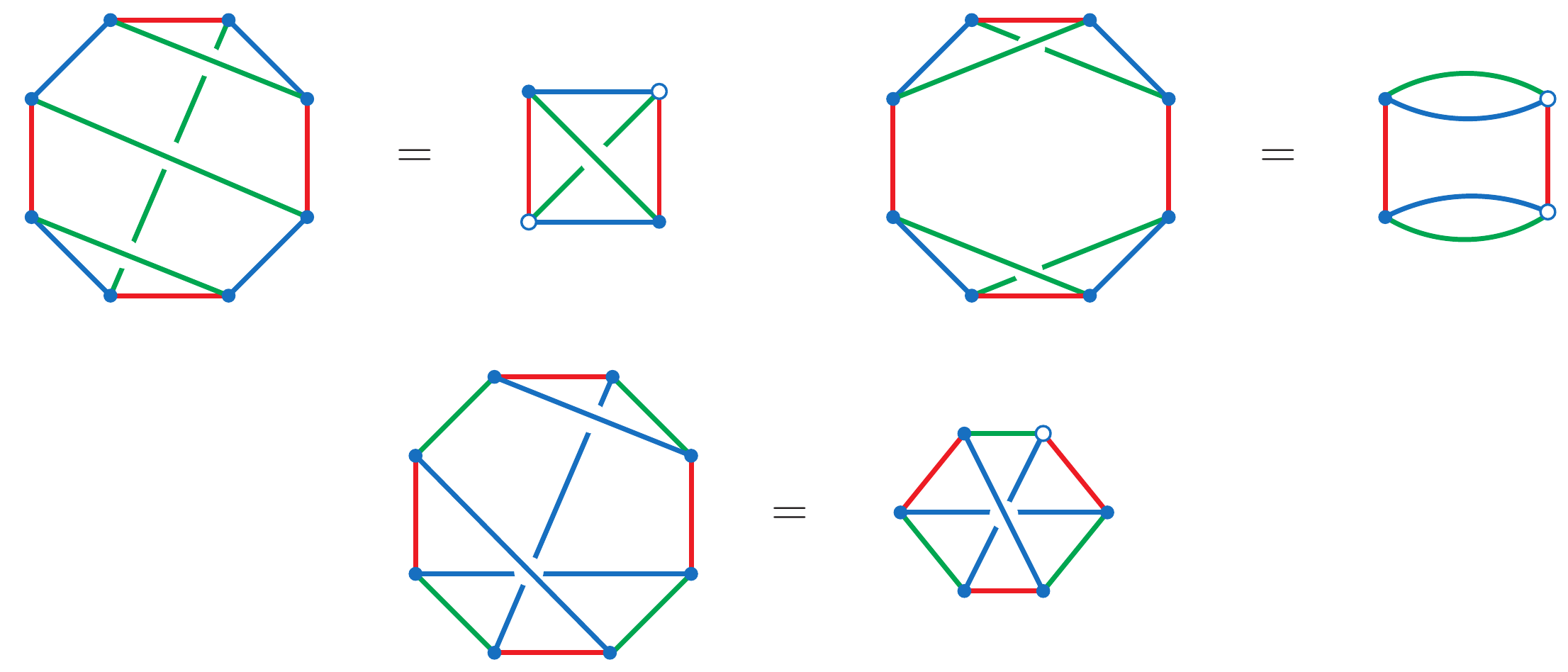}
\caption{The operators which can be represented as lower-order operators with derivative insertions shown by white dots.}
    \label{opequality}
\end{figure}

Furthermore, using the equation of motion (\ref{eom_schem}), we can relate many additional singlet operators to operators containing derivatives.
Let us denote a vertex with $\partial_t \psi$ by a white circle.
By the equations of motion, we can relate the operators whose diagram contains triangles with low-order operators containing derivatives.
For example, some of the operators which can be written as lower-order operators with derivatives are shown in figure \ref{opequality}. 

\begin{figure}[h!]
    \centering
    \includegraphics[width=14cm]{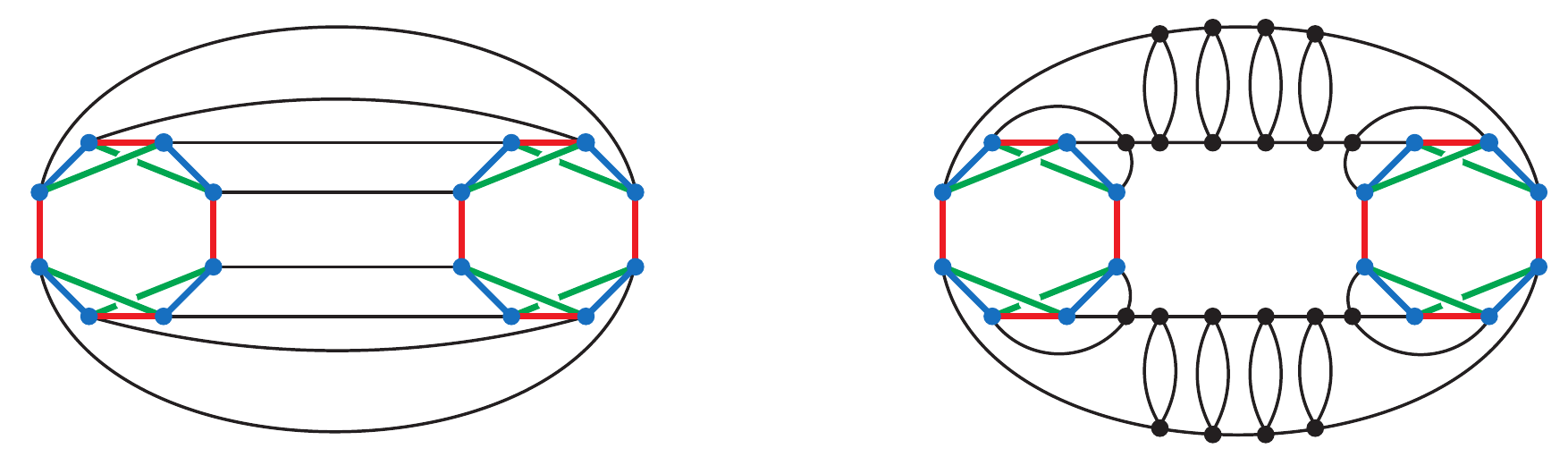}
    \caption{An example of an operator renormalized by two ladder diagrams. The diagram with two ladders inserted (right) is of the same order as the diagram with operators connected directly (left).
    The black dots represent the tetrahedral coupling.}
    \label{fig:O8_2ladder}
\end{figure}

\begin{figure}[h!]
    \centering
    \includegraphics[width=8cm]{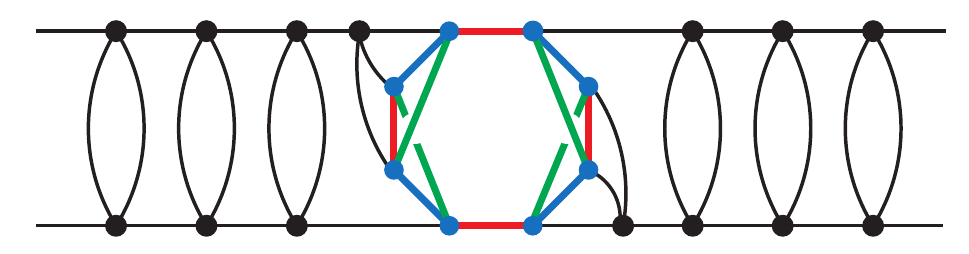}
    \caption{A diagram with two ladders contributing to the correlation function $\langle O_{8}\psi \psi \psi \psi \rangle$.}
    \label{fig:O8_3ladder}
\end{figure}

As discussed in section \ref{composite}, some of these operators are renormalized by multiple ladder diagrams.
For example, the three 4-particle pillow operators, shown in figure \ref{O4ops}, have dimension $h=0$ because they are squares of the symmetry charges. 
Similarly, 
operators ${\cal O}_0^{(a_1 a_2)} {\cal O}_0^{(a_1 a_2)}$ related by the equation of motion to column 2 of figure \ref{O8all}, are renormalized by double ladders
as shown in  figure \ref{fig:O8_2ladder}. One can also see that the correlation function of this operator 
with four fermionic fields receives a contribution from two ladders as shown in  figure \ref{fig:O8_3ladder}

\begin{figure}[h!]
    \centering
    \includegraphics[width=9cm]{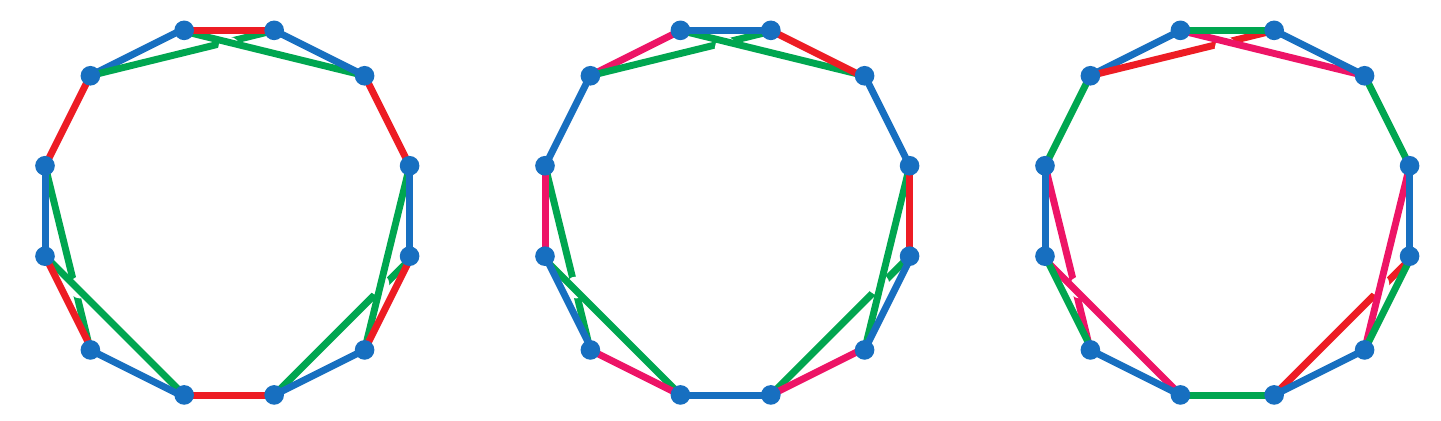}
    \caption{Three 12-particle operators of the same topology, which are renormalized by three-ladder diagrams.}
    \label{fig:O12_3ladder}
\end{figure}

More generally, we may use operators ${\cal O}_n^{(a_1 a_2)}$ defined in (\ref{nonsinglad}) to write down the singlet operators 
\begin{equation}
{\cal O}_{n_1 n_2}=
{\cal O}_{n_1}^{(a_1 a_2)} {\cal O}_{n_2}^{(a_1 a_2)} 
\end{equation}
renormalized by double ladders, 
\begin{equation}
{\cal O}_{n_1 n_2 n_3}=
{\cal O}_{n_1}^{(a_1 a_2)} {\cal O}_{n_2}^{(a_2 a_3)} {\cal O}_{n_3}^{(a_3 a_1)} 
\end{equation}
 renormalized by
triple ladders, and so on. It appears that in the large $N$ limit their scaling dimensions are additive, so that the spectrum of 
${\cal O}_{n_1 n_2}$ is $h_1+h_2$, the spectrum of  ${\cal O}_{n_1 n_2 n_3}$ is $h_1+h_2+h_3$, etc., but we postpone a detailed study
of the relevant Schwinger-Dyson equations. Here $h_i$ are the eigenvalues which appear in the SYK spectrum;
they are the solutions of (\ref{gantisym}).
The picture of the 12-particle operator which is equivalent by the equation of motion to
${\cal O}_{0}^{(a_1 a_2)} {\cal O}_{0}^{(a_2 a_3)} {\cal O}_{0}^{(a_3 a_1)} $, as well as the 
analogous operators ${\cal O}_{0}^{(b_1 b_2)} {\cal O}_{0}^{(b_2 b_3)} {\cal O}_{0}^{(b_3 b_1)} $ and ${\cal O}_{0}^{(c_1 c_2)} {\cal O}_{0}^{(c_2 c_3)} {\cal O}_{0}^{(c_3 c_1)} $, are shown in figure \ref{fig:O12_3ladder}.

\begin{figure}[h!]
    \centering
    \includegraphics[width=8cm]{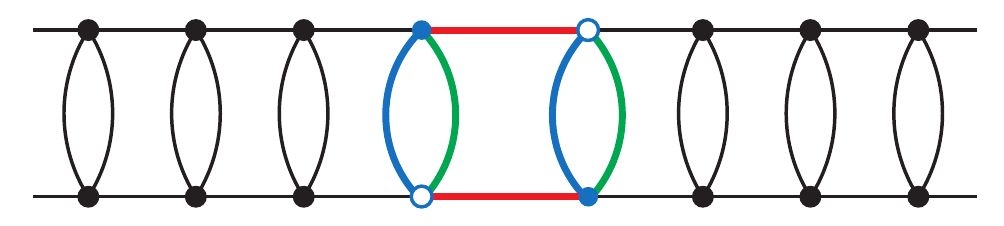}
    \caption{Another representation for the same diagram.}
    \label{PillowLadder}
\end{figure}

We may construct additional operators renormalized by multiple ladders using the operators ${\cal O}_{n}^{[a_1 a_2]}$ (see \ref{nonsinganti}) in addition to ${\cal O}_{n}^{(a_1 a_2)}$. For example,
there is a class of operators ${\cal O}_{n_1}^{[a_1 a_2]} {\cal O}_{n_2}^{[a_1 a_2]}$ whose scaling dimensions appear to be $h_1+ h_2$, where $h_i$ are the solutions of
(\ref{gsym}).
Thus, the charges (\ref{threecharges}) and their products are not the only exceptions to the $m/4$ rule (since the charges are conserved, we a priori expect their scaling
dimension to be zero). In fact, any operator whose diagram contains a bubble subdiagram (i.e. two tensors with a double index contraction) is renormalized by a ladder, and
there are as many ladders as there are bubbles.
For example, a pillow operator contains two bubbles and is renormalized by two ladders.

Moreover, if we take an operator diagram renormalized by multiple ladders and change one vertex in the diagram from $\psi$ to $\partial_t \psi$ (blue to white vertex), it will still be renormalized by the same number of ladders.
With derivatives we can convert a pillow operator into the second operator in fig. \ref{O8n1}.
It is easy to check that this operator is renormalized by two ladders. Since each of the ladders contains the $h=2$ zero-mode in its spectrum, and a zero-mode produces a low-temperature
enhancement by a factor of $\beta J$ \cite{Maldacena:2016hyu}, 
we expect the double-ladder to produce an effect of order $(\beta J)^2$. The multi-ladder enhancements by 
$(\beta J)^n$ seem to be a new effect in the tensor model, which clearly needs to be studied in more detail.

To summarize, we find that:
\begin{enumerate}
    \item The operators containing bubble subgraphs are renormalized with as many ladder diagrams as there are bubble insertions.
    \item The operators obtained from operators with bubble subgraphs by inserting derivatives are renormalized by as many ladders as there were bubble insertions in the original diagram.
    \item The dimensions of operators which are renormalized with a single ladder are given by the solutions of the conformal kernel equation $g\left( h \right)=1$.
    \item The dimensions of the operators which are not renormalized by ladders are multiples of $1/4$.
\end{enumerate}
These results are still far from providing the full information about the singlet spectrum of the $O(N)^3$ tensor quantum mechanics.
In particular, we would like to have a more complete understanding of the operators renormalized by multiple ladders and to study their low-temperature contributions.
We hope to address these questions elsewhere.

\section{Some Scaling Dimensions in the Gurau-Witten Model}

Let us now consider the $O(N)^6$ symmetric quantum mechanical model \cite{Witten:2016iux}. It contains four fermionic
rank-3 tensors $\psi_A$, $A=0,\ldots 3$, each one transforming in the tri-fundamental representation under a different subset of the six $O(N)$ groups.
The four fermionic tensors and the six $O(N)$ gauge groups 
may be visualized as the vertices and edges of a tetrahedron \cite{Witten:2016iux}. Thus, only two of the fermions transform under a given $O(N)$ symmetry.
The Gurau-Witten Hamiltonian is
 \begin{align}
H_{\rm GW}=- \frac{1}{4}g \psi_0^{a b c }\psi_1^{a d e }\psi_2^{f b e}\psi_3^{f d c}\ .
\label{GWHamilt}
\end{align}

The model contains bilinear operators of the form 
$O_A^{c_1 c_2}= \psi_A^{ab c_1} \psi_A^{ab c_2}$.
Let us focus on the operators with $A=0$ and $1$, which transform in the antisymmetric representation of the same $O(N)$ group and can mix with each other: 
\begin{align}
& O_+^{c_1 c_2}= \psi_0^{ab c_1} \psi_0^{ab c_2} + \psi_1^{de c_1} \psi_1^{de c_2}\ , \\
& O_-^{c_1 c_2}= \psi_0^{ab c_1} \psi_0^{ab c_2} - \psi_1^{de c_1} \psi_1^{de c_2}\ .
\label{opmixtures}
\end{align}
The operator $O_+^{c_1 c_2}$ is the charge of one of the six $O(N)$ symmetries; therefore, its scaling dimension vanishes.
The operator $O_-^{c_1 c_2}$ has another scaling dimension, $h_-$. The ladder diagrams contribute to the two-point function
$\langle O_-^{c_1 c_2} (t_1) O_-^{c_3 c_4} (t_2)\rangle $ and we need to derive an appropriate Schwinger-Dyson equation.
If we use $\psi_0^{ab c_1} \psi_0^{ab c_2}$ and $\psi_1^{de c_1} \psi_1^{de c_2}$ as the basis, then
the kernel is a $2\times 2$ symmetric matrix with zeros on the diagonal; hence, the two eigenvalues are equal and opposite. 
To fix the normalization, we note that
the two functions $g_\pm (h)$ are proportional to $\tilde g (h)$, which is given in (\ref{gsym}). 
Therefore, $g_+(h) = \tilde g (h)$ and $g_-(h) = - \tilde g (h)$. The spectrum of solutions to $g_+(h)=1$ indeed includes
$h=0$ corresponding to the conserved charge. The lowest solution to  $g_-(h)=1$ is $h_-\approx 2.33$; this is the scaling dimension of
operator $O_-^{c_1 c_2}$.
Thus, there are three quartic ``pillow operators" made out of $\psi_0$ and $\psi_1$:
$O_+^{c_1 c_2} O_+^{c_1 c_2}$ of dimension $0$, $O_+^{c_1 c_2} O_-^{c_1 c_2}$ of dimension $h_-$, and $O_-^{c_1 c_2} O_-^{c_1 c_2}$ of dimension $2 h_-$.
The third operator is the only pillow operator present in the gauged model where $O_+^{c_1 c_2}$ is set to zero. Its dimension $2h_-\approx 4.66$ makes it very irrelevant;
we find 6 pillow operators with this dimension, corresponding to the presence of 6 different $O(N)$ groups. 

\begin{figure}[h!]
                \centering
                \includegraphics[width=10cm]{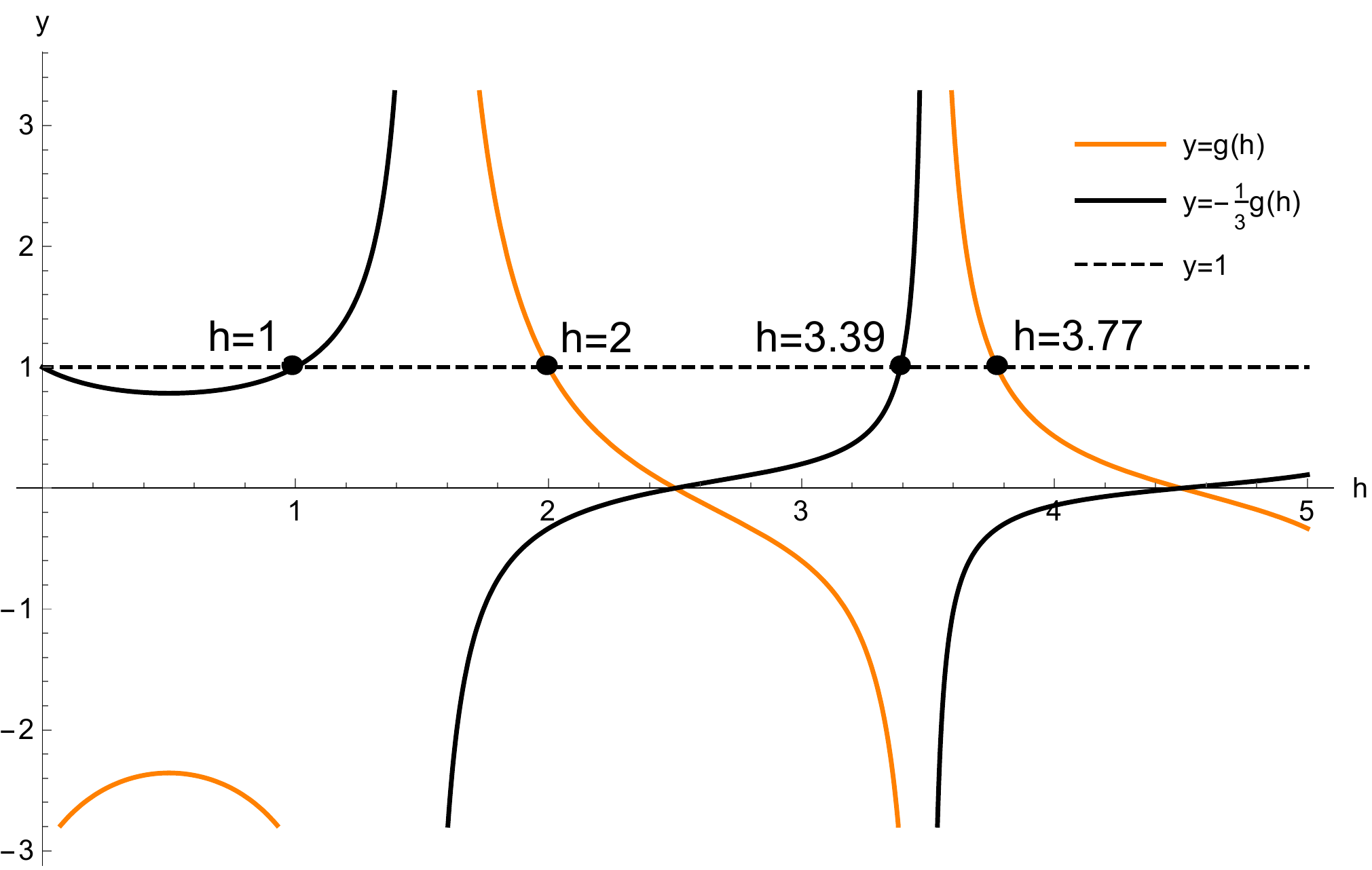}
                \caption{ Plot of the IR dimensions of the bilinear singlet operators in the GW model. }
                \label{GWplot}
\end{figure}

We may also study the bilinear singlet operators like
\begin{align}
O_-^n=
\psi_0^{ab c} \partial_t^{2n+1} \psi_0^{ab c} - \psi_1^{de c} \partial_t^{2n+1}\psi_1^{de c}\ .
\end{align}
For $n=0$ this operator vanishes after the use of equations of motion, but it is non-trivial for $n=1, 2, \ldots$.  
To calculate the scaling dimensions of these operators using the S-D equations we note that the kernel is the SYK kernel,
\begin{align}
K_{\rm SYK} (t_{1},t_{2};t_{3},t_{4})  =
-\frac{3} {4\pi} \frac{\sgn(t_{1}-t_{3}) \sgn(t_{2}-t_{4})  } {|t_{1}-t_{3}|^{1/2}|t_{2}-t_{4}|^{1/2}|t_{3}-t_{4}|}
 \, ,
\end{align}
times a  $4\times 4$ matrix with zeros on the diagonal, and all the
off-diagonal elements equal to the same value $B$. To determine $B$, we note that the kernel corresponding to the eigenvector $(1,1,1,1)$ with eigenvalue $3B$
should exactly equal the SYK kernel. This means that $B=1/3$, which gives the spectrum of the SYK model determined by $g (h)=1$ (see \ref{gantisym}).
The three eigenvectors $(1,-1,0,0)$, $(0,1,-1,0)$, $(0,0,1,-1)$ have eigenvalue $-B=-1/3$; thus, the spectrum of corresponding operators is determined by 
\begin{align}
-{1\over 3}
g (h)=1\ .
\label{triplets}
\end{align}
The solutions to this equation are shown in figure \ref{GWplot}.\footnote{
We may decompose the
$O(N)^6$ invariant operators into irreducible representations of the symmetry group of the tetrahedron, which is isomorphic to $S_4$. Each solution
to (\ref{triplets}) corresponds to 3 operators belonging to a degree $3$ representation of $S_4$. 
} 
There is a series of solutions that lie slightly below $2n+\frac 3 2$, for $n=1, 2, 3, \ldots$ and approach it at large $n$. 
In other words, they lie slightly below the naive dimensions of operators $O_-^n$. 
For $n=1$ the numerical value is $3.39$, which is close to $3.5$.
There is also an exact solution with $h=1$, whose interpretation is not completely clear.

The dimensions of operators  $O_-^n$ that we find are the same as in the Gross--Rosenhaus ``generalized SYK model" 
 \cite{Gross:2016kjj} for $q=4$. In particular, the $h=1$ solution is present in that case as well, and the corresponding operator decouples.
The Gross-Rosenhaus model that corresponds to the colored tensor model has $f=4$, i.e. it contains four flavors of Majorana fields, $\chi_a^i$, $a=1, \ldots, 4$. Its Hamiltonian may be written as 
\begin{align}
H= J_{ijkl} \chi_i^1 \chi_j^2 \chi_k^3 \chi_l^4   
\ ,\end{align}
where $J_{ijkl}$ are random couplings.
The operators which are analogous to $O_-^n$ are $\chi_i^1 \partial_t^{n+1} \chi_i^1- \chi_j^2 \partial_t^{n+1} \chi_j^2 $.
The $n=0$ operator vanishes by the equation of motion for any value of $J_{ijkl}$, which appears to explain the decoupling
of the $h=1$ mode.

\section{Counting singlet operators in $d=1$}

In this section we proceed to do the singlet operator counting in the $O(N)^3$ quantum mechanics more systematically.
We employ the technique used in \cite{Aharony:2003sx, Beccaria:2017aqc} to find the partition function and free energy of gauge theory. 
In our case, we will see that the free energy diverges wildly, but nevertheless this procedure allows to count the operators in the gauged or ungauged fermionic and scalar theories.

We work in the one-dimensional spacetime with fields living in the tri-fundamental representation of $O(N)_1 \times O(N)_2 \times O(N)_3$, in the limit of $N \to \infty$. 
We will mainly address the case of the free tensor model, which describes the UV fixed point, but also make comments about the IR theory.
The partition function may be written in the form:
\begin{equation}
    \mathcal Z=\sum_{\mathcal O_i}x^{h_i}, \qquad x\equiv e^{-\beta},
    \label{Z_def}
\end{equation}
where $\mathcal O_i$ are all operators in the theory which are singlets under $O(N)^3$.
Here $h_i$ are the conformal dimensions, so in the UV this partition function is
\begin{equation}
    \mathcal Z = \sum_k n_k x^{k h_{UV}},
    \label{Z_UV}
\end{equation}
where $k$ is the number of fields comprising an operator and $n_k$ is the number of admissible operators for each $k$.
In what follows we call $k$ the order of an operator. For the fermionic model $h_{UV}= (d-1)/2$, and for bosonic it is $(d-2)/2$. 

The partition function counts all operators including the disconnected ones. 
To restrict ourselves exclusively to the connected operators, we have to compute the single-sum partition function defined as:
\begin{equation}
    \log \mathcal Z(x)=\sum_{m=1}^\infty \frac{1}{m}\mathcal Z_{\text{s.s.}}\left( x^m \right).
    \label{Z_st_def}
\end{equation}
To find $\mathcal Z_{\text{s.s.}}$ explicitly, we use an elegant formula from \cite{Beccaria:2017aqc}:
\begin{equation}
    \mathcal Z_{\text{s.s.}}\left( x \right)= \log \mathcal Z(x)+\sum_{m \in \Omega}\left( -1 \right)^{\nu_m}\frac{1}{m}\log \mathcal Z\left( x^{m} \right).
    \label{Z_st_form}
\end{equation}
Here $m$ belongs to the set of square-free integers $\Omega=\{2, 3, 5, 6, 7, 10, 11, 13,\dots\} $:
\begin{equation}
    m=\prod_{i=1}^{\nu_m}p_i, \qquad p_i \text{ prime}\,.
    \label{m_def}
\end{equation}

Our goal in this section is to find the single-sum partition function for the scalar and fermionic tensor models.
The partition function for the scalar theory in the UV with one group can be found as \cite{Sundborg:1999ue,Polyakov:2001af,Aharony:2003sx}:
\begin{equation}
    \mathcal Z^S=\int dM \exp\left( \sum_{m=1}^\infty \frac{1}{m} z_{S,d}(x^m) \chi(M^m) \right),
    \label{Z_S_def}
\end{equation}
and for the fermionic theory it is:
\begin{equation}
    \mathcal Z^F=\int dM \exp\left( \sum_{m=1}^\infty \frac{(-1)^{m+1}}{m} z_{F,d}(x^m) \chi(M^m) \right),
    \label{Z_F_def}
\end{equation}
with $M$ in the symmetry group and $\chi(M)$ being the character of the desired representation.
In our case, we substitute:
\begin{equation}
    M \to M_1 M_2 M_3, \qquad \chi(M) \to \chi(M_1) \chi(M_2) \chi(M_3), \qquad M_i \in O(N)_i \, 
    \label{M_M123}
\end{equation}
and take $\chi(M)=\tr M$.

The single-letter partition functions for scalars and (Majorana) fermions correspondingly are as follows:
\begin{equation}
    z_{S,d}(x)=\frac{x^{\frac{d}{2}-1}(1+x)}{(1-x)^{d-1}}\,,
    \label{z_s}
\end{equation}
\begin{equation}
    z_{F,d}(x)=\frac{2^{\lfloor \frac{d}{2}\rfloor}x^{\frac{d-1}{2}}}{(1-x)^{d-1}} \,.
    \label{z_f}
\end{equation}
To find $\mathcal Z$, we will need the integrals of characters of $O(N)$ \cite{Beccaria:2017aqc}:
\begin{equation}
    \int dM \prod_{l} \left( \tr M^l \right)^{a_l}=\prod_l \left[\begin{matrix}
            l \text{ odd, } a_l \text{ even} & \left( 2l \right)^{a_l/2} \frac{1}{\sqrt{\pi}} \Gamma\left( \frac{a_l}{2}+\frac12 \right),\\
            l \text{ even} & \sum_{k=0}^{a_l/2} { {a_l} \choose {2k} } \left( 2l \right)^k \frac{1}{\sqrt{\pi}} \Gamma\left( k+\frac12 \right).
        \end{matrix}\right.
    \label{I_calc}
\end{equation}

In the next chapter, we first find partition functions for both the fermionic and scalar $d=1$ models without the constraint that the charges
(\ref{threecharges}) vanish.
Then, to find the partition function for the operators in the gauged model, we subtract the contribution 
from the operators containing $O(N)$ charge, or a ``bubble'' subdiagram  (\ref{threecharges}) (see fig. \ref{Charges}).
Such operators should vanish in the gauged version of quantum mechanics.

\subsection{Fermions}

The single-letter partition function for real fermions $z_{F,d}$ is not well defined in one dimension. 
This reflects the divergence of the partition function (and hence free energy).
To regularize it, we formally proceed in $\left( 1+2\epsilon \right)$ dimension and neglect all the terms proportional to $\epsilon$ in the single-letter partition function; in other words, we simply take:
\begin{equation}
    z_{F,1+2\epsilon}=x^\epsilon\,.
    \label{z_f_1+e}
\end{equation}

We can justify this choice as follows. 
The single-letter partition function counts all local operators containing one field $\psi^{abc}$ with any number of derivatives. 
In our case, the only such operator is $\psi^{abc}$: since $\partial_t \psi^{abc}$ vanishes by equations of motion in the free theory, 
all the operators with higher derivatives will vanish too.

In other words, in the fermionic case we are counting only the operators made of fermions without derivatives.
We can think of this as operator counting in a $d=0$ model (for a review see \cite{Gurau:2011xp}), but with the Fermi statistics imposed.

Computing $\mathcal Z$ and using (\ref{Z_F_def}), (\ref{I_calc}), we find to first several orders in $x$:
\begin{equation}
    \mathcal Z^F=1 + 4 x^{4\epsilon} + 70 x^{8\epsilon} + 116 x^{10 \epsilon} + 3062 x^{12 \epsilon} + 24788 x^{14\epsilon}+409869 x^{16 \epsilon}+\dots\,.
    \label{Z_F_ans}
\end{equation}
From this we can find the single-sum partition function, which counts connected operators:
\begin{equation}
    \mathcal Z_{\text{s.s.}}^F=4 x^{4\epsilon} + 60 x^{8\epsilon} + 116 x^{10\epsilon} + 2802 x^{12\epsilon} + 24324 x^{14 \epsilon}+396196 x^{16\epsilon}+\dots\,.
    \label{Z_st_F_ans}
\end{equation}
The order $2k$ in $x^{2k\epsilon}$ gives the number of fermions in the operator.
So we see there are four four-fermion operators: one tetrahedron and three differently colored pillows (see figure \ref{O4ops}).
Note that, although we employed a gauged theory to count these operators, the pillows and other operators containing $O(N)$ charges are still present.
At the sixth order, there are no operators because of the Fermi statistics as we noticed before, but at order $8$ there are 60 operators.

\begin{figure}
    \begin{tabular}{cc}
    \includegraphics[width=.5\textwidth]{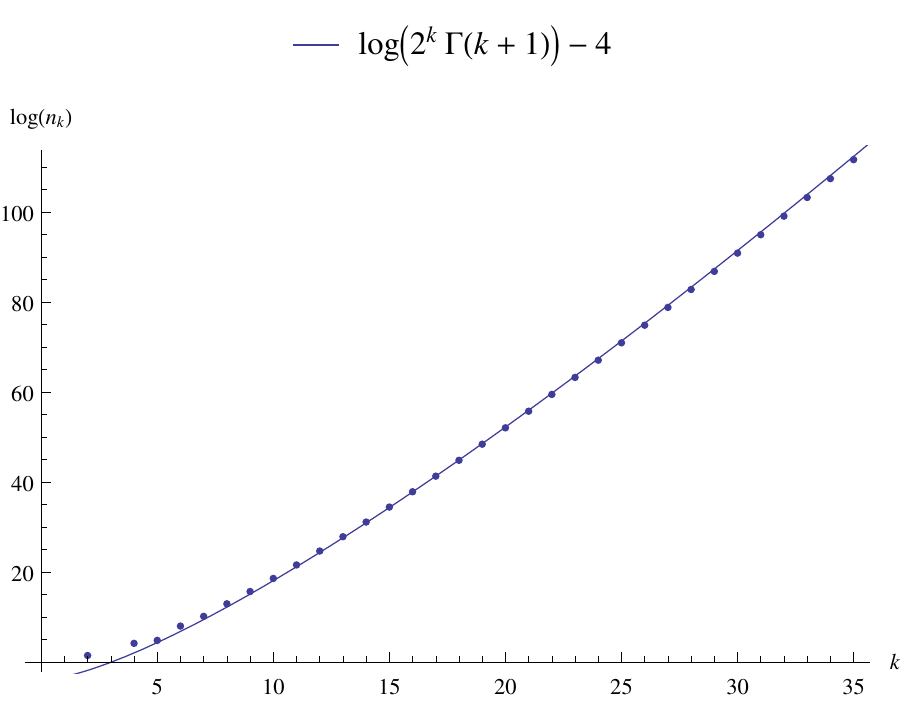}&         
    \includegraphics[width=.5\textwidth]{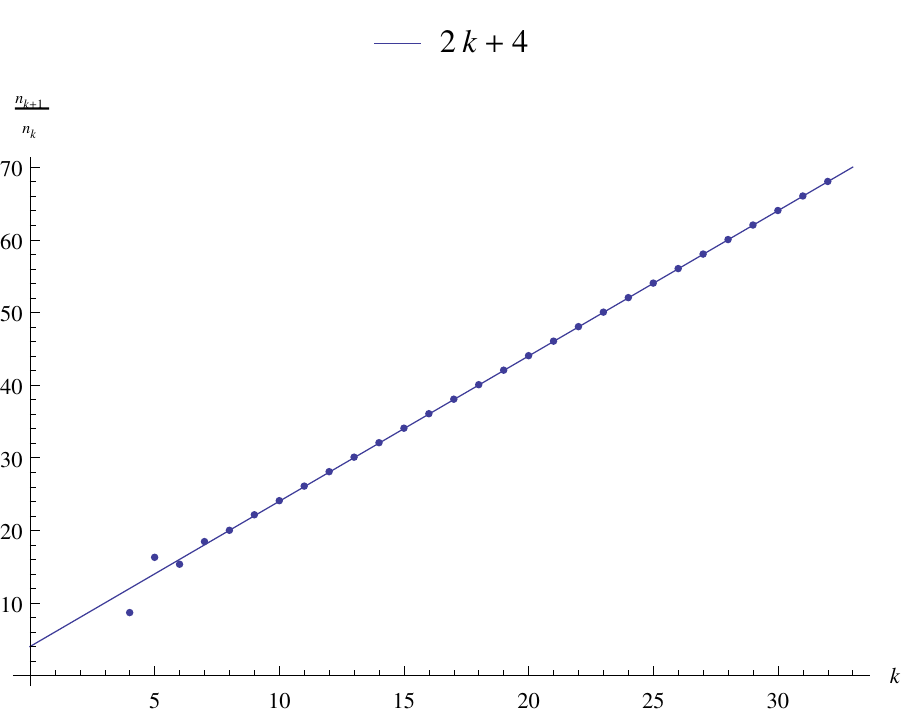}         
    \end{tabular}
    \caption{Logarithm of the number of allowed $(2k)$-particle fermionic operators as a function of $k$. We see that the number of operators grows like $\sim k! 2^k$.}
    \label{fig:oper_mult}
\end{figure}

The number of $2k$-particle operators grows roughly as (see fig. \ref{fig:oper_mult}):
\begin{equation}
    n_{2k}\sim 2^k k!  
    \label{n(k)_ferm_ungauged}
\end{equation}

\begin{figure}
    \centering
    \includegraphics[width=.45\textwidth]{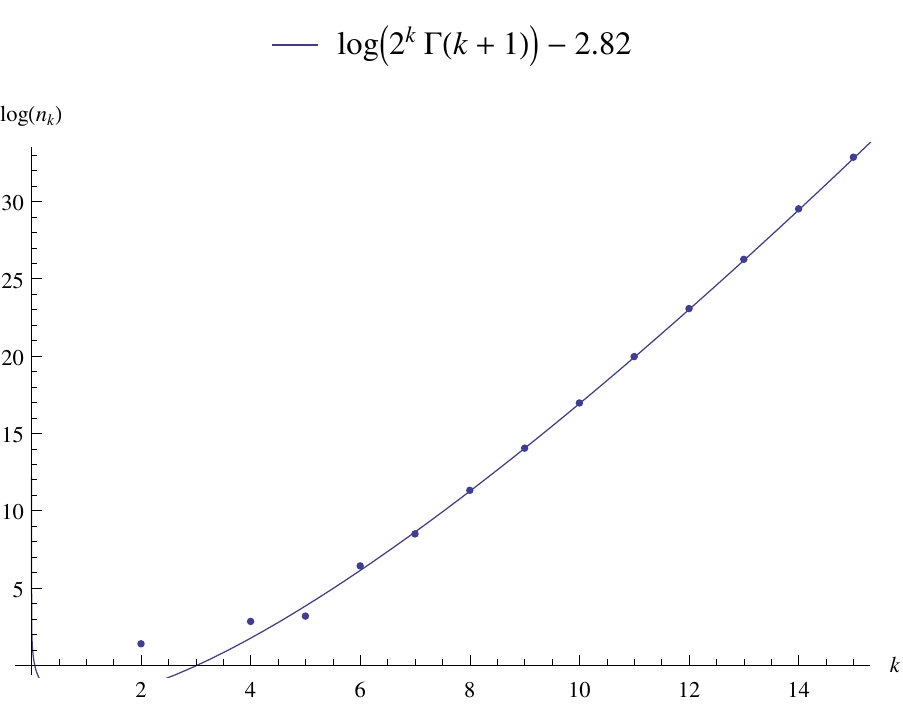}
    \includegraphics[width=.45\textwidth]{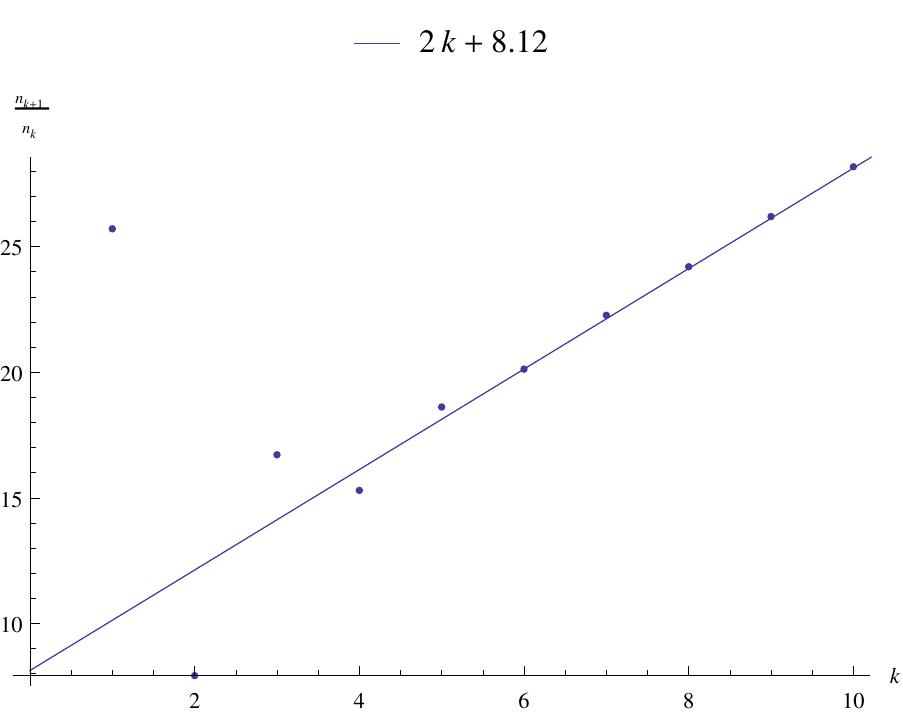}
    \caption{Left: The logarithm of the number of $(2k)$-particle operators $n_{2k}$ in the model where $O(N)^3$ symmetry is gauged. 
The asymptotic of the number of operators is roughly the same as in the ungauged theory. 
Right: the ratio $n_{2k+2}/n_{2k}$ plotted against $k$. The linear
behaviour clearly indicates $\sim 2^k k!$ growth. }
    \label{fig:mult_gauged}
\end{figure}

To count operators in the gauged model where the vanishing of $O(N)$ charges (\ref{threecharges}) is imposed, we have to disregard the operators containing their insertions, i.e. the ``bubble'' subgraphs.
In order to do that, we subtract the operators having the same quantum numbers as a bubble in the exponent of (\ref{Z_F_def}). Each $O(N)$ charge 
(\ref{threecharges}) is antisymmetric in its two indices,
which means that it lives in the representation $\left( N \otimes N \right)_{\text{antisym}}$ with the character:
\begin{equation}
    \chi_A\left( M \right)\equiv\chi_{\left( N \otimes N \right)_{\text{antisym}}}\left( M \right)=\frac12 \left( \left( \tr M \right)^2 - \tr M^2  \right).
    \label{chi_antisym}
\end{equation}
The bubble is a bosonic operator and its conformal dimension in the UV is $2\epsilon$.
Bringing it all together, we find that the partition function for operators in the gauge theory is:
\begin{align}
    \mathcal Z^{F\text{(gauge)}}=&\int dM_{1}dM_{2}dM_{3}\exp\bigg( \sum_{m=1}^\infty \frac{1}{m}\Big( (-1)^{m+1}x^{m\epsilon} \chi(M_1)\chi\big( M_2 \big)\chi\big( M_3 \big)\notag\\
    &-x^{2m\epsilon}\big( \chi_A\big( M_1 \big)+\chi_A\big( M_2 \big)+\chi_A\big( M_3 \big)  \big)\Big)\bigg).
    \label{Z_F_gauge}
\end{align}
The single-sum partition function for the gauge theory then is as follows:
\begin{equation}
    \mathcal{Z}^{F\text{ (gauge)}}_{\text{s.s.}}= x^{4\epsilon} + 17 x^{8\epsilon} + 24 x^{10\epsilon} + 617 x^{12\epsilon} + 4887 x^{14 \epsilon} + 82466 x^{16\epsilon}+\dots\,.
    \label{Z_F_gauge_ans}
\end{equation}
We see that at the fourth order we are left with one operator; namely, the tetrahedron.
At the eighth order we see 17 operators, as we already found in section \ref{eightpart} via explicit construction (see fig. \ref{O8all})
We have computed the single-sum partition function up to order 30, and the result matches the same 
factorial growth as in the model where the $O(N)^3$ symmetry is not gauged (see fig. \ref{fig:mult_gauged}). 

Finally, let us comment on the IR theory, where
we believe there is similarly rapid growth of the number of operators as
a function of the conformal dimension. Since for the majority of $2k$-particle operators the large $N$ IR dimension is $h=k/2$, in view of   
the result (\ref{n(k)_ferm_ungauged}) we expect that the number of operators of dimension $h$ to grow as $\Gamma(2h+1)$, up to an exponential prefactor.

\subsection{Bosons}

We can also count the allowed operators in the scalar theory. 
Proceeding in the same fashion, we define single-letter partition function in $(1+2\epsilon)$ dimensions as follows:
\begin{equation}
    z_{S,1+2\eps} = x^{-\frac12+\epsilon}(1+x)\,,
    \label{z_s_1}
\end{equation}
where $-\frac12+\epsilon$ is the dimension of the scalar field.
The partition function is:
\begin{multline}
    \mathcal Z^S=1+x^{2\epsilon}\left( x^{-1}+1+x \right)+x^{4\epsilon}\left( 5x^{-2}+5x^{-1}+14+5x+5x^2 \right)\\
    +x^{6\epsilon}\left( 16 x^{-3}+34 x^{-2}+ 101 x^{-1}+108+101x+34x^2+16x^3 \right)+\dots\,.
    \label{Z_S_ans}
\end{multline}
The single-sum partition function, which includes the operators with bubble insertions, is:
\begin{multline}
    \mathcal Z_{\text{s.s.}}^S=x^{2\epsilon}\left( x^{-1}+1+x \right)+x^{4\epsilon}\left( 4x^{-2}+4x^{-1}+12+4x+4x^2 \right)\\
    +x^{6\epsilon}\left( 11 x^{-3}+25 x^{-2}+ 79 x^{-1}+86+79x+25x^2+11x^3 \right)+\dots\,.
    \label{Z_st_S_ans}
\end{multline}
In the second order we have operators $\phi^{abc} \phi^{abc}$, $\phi^{abc} \partial_t \phi^{abc}$, and $\partial_t \phi^{abc} \partial_t \phi^{abc}$. 
In the fourth order, we find the pillows and tetrahedra with various insertions of $\partial_t$.
This partition function also diverges at $\epsilon \to 0$ and displays the factorial growth of the number of operators with their order.

To count operators in the gauged theory, we once again have to take care of the subgraphs corresponding to the gauge group charge. 
For a scalar theory, the gauge charge operator is:
\begin{equation}
    Q^{a_1 a_2}=\phi^{a_1 bc}\overleftrightarrow{\partial_{t}}\phi^{a_2 bc}\,.
        \label{Q_sym}
\end{equation}
This operator lives in the adjoint representation, just like the gauge field. 
Its dimension is $2\epsilon=\left( -\frac12 + \epsilon \right)+\left( \frac12 + \epsilon \right)$.
The character of the adjoint representation is:
\begin{equation}
    \chi_{\text{adj}}\left( M \right)=\frac12 \left( \left( \tr M \right)^2-\tr M^2\right).
    \label{chi_S_def}
\end{equation}
Taking all this into account, we write the partition function as:
\begin{multline}
    \mathcal Z^{S\text{(gauge)}}=\int dM_1 dM_2 dM_3 \exp \left( \sum_{m=1}^{\infty} \frac1m \left( \left( x^{-\frac{m}{2}+\epsilon m}+x^{\frac{m}{2}+\epsilon m} \right)\chi\left( M_1 \right)\chi\left( M_2 \right)\chi\left( M_3 \right)\right.\right.\\
    \left.\left.- \chi_{\text{adj}}\left( M_1 \right)x^{2m\epsilon}- \chi_{\text{adj}}\left( M_2 \right)x^{2m\epsilon}- \chi_{\text{adj}}\left( M_3 \right)x^{2m\epsilon}\right)\right).
    \label{Z_S_gauge}
 \end{multline}
To the first six orders, the partition function reads as:
\begin{multline}
    \mathcal Z^{S\text{(gauge)}}=1+x^{2\epsilon}\left( x^{-1}+1+x \right)+x^{4\epsilon}\left( 5x^{-2}+5x^{-1}+11+5x+5x^2 \right)\\
    +x^{6\epsilon}\left( 16 x^{-3}+34 x^{-2}+77 x^{-1}+84+77x+34x^2+16x^3 \right)+\dots\,.
    \label{Z_S_gauge_ans}
\end{multline}
The single-sum partition function, which counts only the operators with connected diagrams, is as follows:
\begin{multline}
    \mathcal Z^{S\text{(gauge)}}_{\text{s.s.}}=x^{2\epsilon}\left( x^{-1}+1+x \right)+x^{4\epsilon}\left( 4x^{-2}+4x^{-1}+9+4x+4x^2 \right)\\
    +x^{6\epsilon}\left( 11 x^{-3}+25 x^{-2}+58 x^{-1}+65+58x+25x^2+11x^3 \right)+\dots\,.
    \label{Z_S_gauge_st}
\end{multline}
The first term in this expression corresponds to the operators $\phi^{abc}\phi^{abc}$, $\phi^{abc}\partial_t\phi^{abc}$, and $\partial_t\phi^{abc}\partial_t\phi^{abc}$
(the second of these operators is a total derivative; such descendant operators are included in the counting). 
The number 11 in the third term corresponds to all the six-particle graphs discussed in 
Section \ref{gaugeinv}. Now the number of operators containing a string of $2k$ scalars is approximately
\begin{equation}
n_{2k} \sim 2^{2k} \times 2^k k! 
\end{equation}
Compared to the fermionic case \ref{n(k)_ferm_ungauged} we have an additional factor of $2^{2k}$. 
As we will see in the next section, for $d=0$ the leading asymptotic for the number of operators is the same for scalars and
fermions. Therefore, the factor $2^{2k}$ comes from distributing the time derivatives $\partial_t$ among $2k$ fields. Since in the free theory 
$\partial_t^2 \phi^{abc}=0$, each of the $2k$ fields may be acted on by one or no derivatives. This indeed contributes a factor of $2^{2k}$.

\section{Counting the Invariants in $d=0$}
\label{countinv}

Here we use methods similar to those in the previous section to discuss the counting of invariants in the $d=0$ model which is simply an integral over
the tensor.
The construction and counting of such invariants, which are made out of products of tensors with all indices contracted,
has been addressed in \cite{Gurau:2011xp,Geloun:2013kta,Itoyama:2017xid,Mironov:2017aqv,Diaz:2017kub,deMelloKoch:2017bvv}.
These papers primarily discuss the complex bosonic rank-$r$ tensor models which possess $U(N)^r$ symmetry.
We will first consider the bosonic rank-$3$ tensor model with $O(N)^3$ symmetry and perform the counting using the methods developed in 
\cite{Aharony:2003sx, Beccaria:2017aqc}. The model of a real fermionic tensor $\psi^{abc}$ does not work in $d=0$: since the $O(N)^3$
invariant $\psi^{abc}\psi^{abc}$ vanishes, it is impossible to write down a Gaussian integral.
One can write down models of complex fermionic tensors in $d=0$, but we won't study them here.
We will address the bosonic rank-$3$ symmetric traceless and antisymmetric tensors in subsection \ref{symmantisymm},
and the bosonic complex tensors with $U(N)^3$ and $U(N)^2 \times O(N)$ symmetries
in subsection \ref{complexten}.

The single-letter partition function counts all the invariants containing one field.
In our case the only such operator is $\phi^{abc}$, so the single-letter partition function is:
\begin{equation}
    z_{S,0}(x)=x\,.
    \label{z_S_0}
\end{equation}
The invariants in this case are given by the diagrams with $2k$ vertices and three edges of different colors meeting at each vertex.
Thus, the invariants are isomorphic to the Feynman diagrams in the theory of three scalar fields with interaction $\varphi_1 \varphi_2 \varphi_3$.
Every edge of the diagram is assigned one of the three colors, and every vertex joins the edges of three different colors.
This is a non-trivial condition; for example, one-particle reducible graphs cannot be colored in this way.
We consider different colorings of the diagrams as different invariants, so each topology can enter multiple times if there are several distinct ways to color it.

Using (\ref{Z_F_def}), we find the full partition function:
\begin{equation}
    \mathcal Z^0=\int dM_1dM_2dM_3 \exp\left( \sum_{m=1}^\infty \frac{1}{m} x^m \chi(M_1^m) \chi\left( M_2^m \right) \chi\left( M_3^m \right) \right),
    \label{Z_S_0}
\end{equation}
where we have used the character of a tri-fundamental representation (\ref{M_M123}).
Taking this integral and using (\ref{I_calc}), we find in the first several orders:
\begin{equation}
    \mathcal Z^0=1 + x^2 + 5 x^4 + 16 x^6 + 86 x^8 + 448 x^{10}+3580 x^{12} + 34981 x^{14}+\dots\,.
    \label{Z_S_0_ans}
\end{equation}
This partition function counts all the invariants, including the disconnected ones.
To remove the latter, we compute the single-sum partition function using (\ref{Z_st_form}):
\begin{equation}
    \mathcal Z_{\text{s.s.}}^0=x^2 + 4 x^4 + 11 x^6 + 60 x^8 + 318 x^{10}+2806 x^{12} + 29359 x^{14}+\dots\,.
    \label{Z_S_0_st_ans}
\end{equation}
The only two-scalar invariant is $\phi^{abc}\phi^{abc}$. 
The four four-scalar invariants are the three inequivalent pillows and the tetrahedron, shown in figure \ref{O4ops}.
The eleven six-scalar invariants are the ones shown in fig. \ref{O6ops}.

\begin{figure}
    \begin{tabular}{cc}
        \includegraphics[width=.5\textwidth]{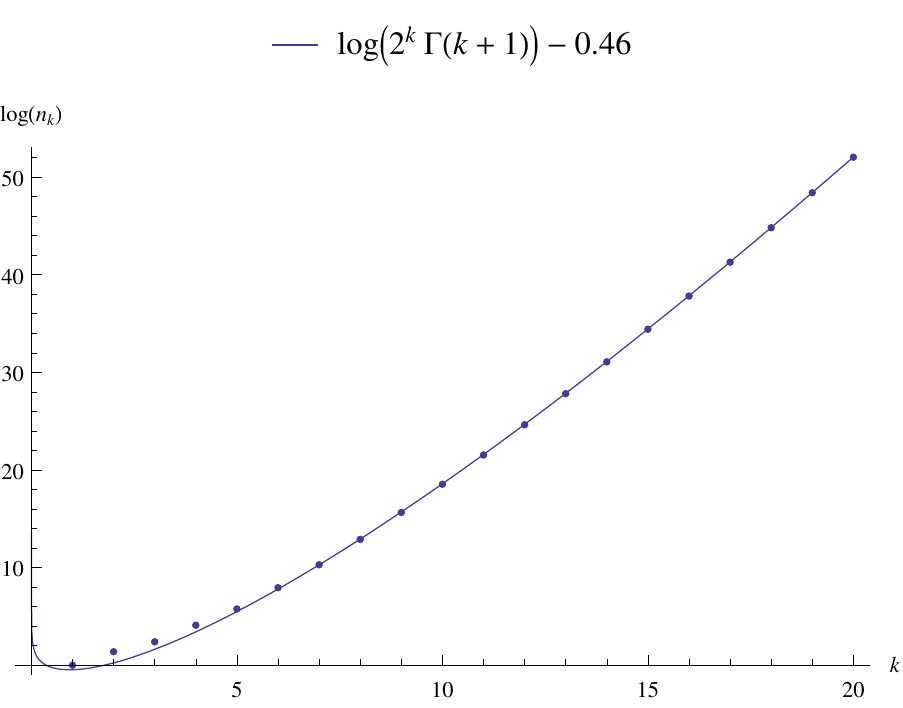}&
        \includegraphics[width=.5\textwidth]{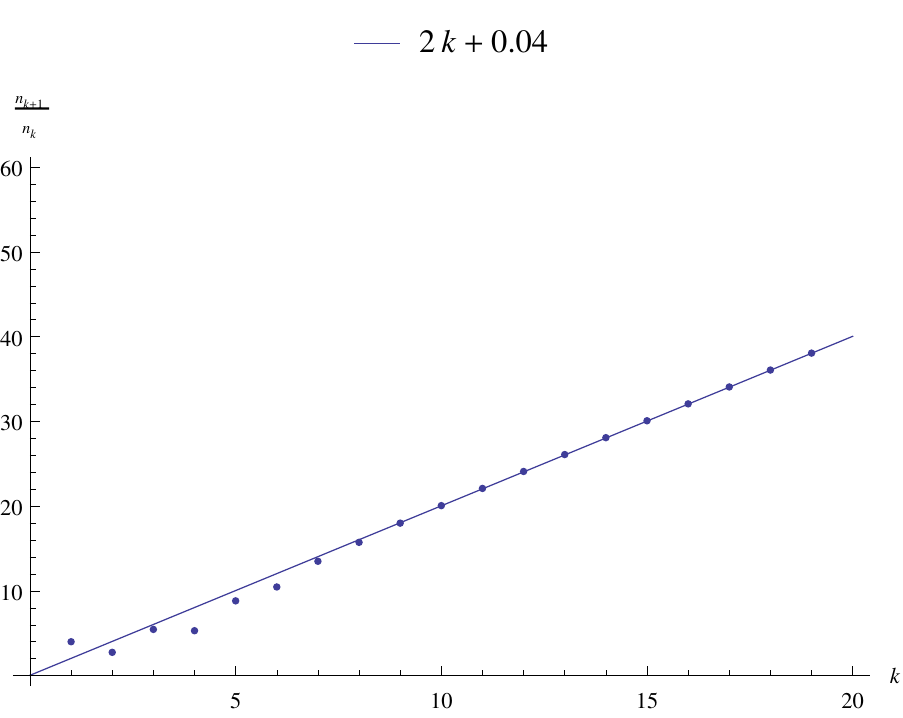}
    \end{tabular}
    \caption{Logarithm of the number of invariants with $2k$ scalars as a function of $k$. The number grows as $\sim k! 2^k$.}
    \label{fig:mult_d0}
\end{figure}

The number of invariants made out of $2k$ fields grows asymptotically as (see fig.\ref{fig:mult_d0}):
\begin{equation}
    n_{2k} \sim 2^k k! 
    \label{n_k_0}
\end{equation}
We can find this asymptotic from an analytic estimate. The key observation is that the integral (\ref{I_calc}) grows factorially as $(a_l/2)!$ for large $a_l$, while only as a power
$l^{a_l/2}$ for large $l$. Besides, for large $a_l$ there is no difference in the leading order between odd and even $l$. 
Therefore, the leading contribution to $x^{2k}$ will come simply from the $m=1$ term:
\begin{equation}
\label{analytic}
n_{2k}\sim \cfrac{1}{(2k)!} \int \ dM_1 dM_2 dM_3 \ (\chi(M_1) \chi(M_2) \chi(M_3))^{2k} = \cfrac{1}{(2k)!} 
\left( 2^{k} \Gamma \left( k+1/2 \right) \right)^3 \sim 2^k k!
\end{equation}
Since the dominant term originates only from $m=1$ term, the same estimate is valid for the fermions.

\subsection{Symmetric traceless and antisymmetric tensors}
\label{symmantisymm}

Let us also discuss the counting of invariants in
models with a single $O(N)$ symmetry, where we will consider the tensors which are either symmetric traceless or fully antisymmetric. Such models with the tetrahedral interactions were
recently studied in \cite{Klebanov:2017nlk}, where evidence was provided that they have melonic large $N$ limits. The full partition function is
\begin{equation}
    \mathcal Z=\int dM  \exp\left( \sum_{m=1}^\infty \frac{1}{m} x^m \chi(M^m)  \right),
    \label{Z_ON}
\end{equation}
where for the 3-index symmetric traceless representation the character in the large $N$ limit is \footnote{The 
more complicated expression at finite $N$ may be extracted from eq. (2.4) of \cite{Klebanov:2017nlk}.}
\begin{align}
\chi^+(M) = \frac 1 6  (\tr M)^3 + \frac 1 2 \tr M \tr M^2 + \frac 1 3 \tr M^3- \tr M
\ .
\label{symchar}\end{align}
For the fully antisymmetric representation the character is
\begin{align}
\chi^- (M) = \frac 1 6  (\tr M)^3 - \frac 1 2 \tr M \tr M^2 + \frac 1 3 \tr M^3
\ .\end{align}

In the symmetric traceless case, the partition function is found to be
\begin{equation}
    {\mathcal Z}^+ =1+ x^2 +3 x^4 + 9 x^6 + 32 x^8 + 135 x^{10}+ 709 x^{12} +\dots\,.
    \label{Z_symmtr}
\end{equation}
Extracting the single-sum expression, we find
\begin{equation}
    \mathcal Z_{\text{s.s.}}^+ =x^2 + 2 x^4 + 6 x^6+20x^8+91x^{10}+509x^{12}+ \dots\,.
    \label{Z_symmtrss}
\end{equation}
The numbers of $O(N)$ invariants made of $2k$ fields are the same as the numbers of connected tadpole-free vacuum diagrams in the $\phi^3$ theory
(here the edges have only one color). 
They are smaller than the corresponding numbers in (\ref{Z_S_0_st_ans}) referring to the $O(N)^3$ theory. 
For example, at order $4$ 
we now have only $2$ distinct invariants: in addition to the tetrahedron there is only one pillow, since there are no distinct colorings of it. 
For large $k$ the number of invariants
can be estimated similarly to the tri-fundamental case (\ref{analytic}). Once again, the term with $m=1$ dominates. Moreover,
out of the four terms in (\ref{symchar}), $(\tr M)^3/6$ gives the biggest contribution. Therefore,
\be
n^{\pm}_{2k} \sim \cfrac{1}{(2k)! 6^{2k}} \int dM (\tr M)^{6k}  \sim \left( \frac 3 2 \right)^k k!
\ee
where we used the integrals (\ref{I_calc}). 

Since $(\tr M)^3/6$ dominates, the same asymptotic formula is valid for the 
3-index antisymmetric case.
Here the partition function is found to be
\begin{equation}
    {\mathcal Z}^- =1+ x^2 +3 x^4 + 7 x^6 + 24 x^8 + 86x^{10}+426x^{12} +\dots\ ,
    \label{Z_antisymm}
\end{equation}
and the single-sum partition function is 
\begin{equation}
    \mathcal Z_{\text{s.s.}}^- =x^2 + 2 x^4 + 4 x^6+14x^8+54x^{10}+298x^{12}+ \dots\,.
    \label{Z_antisymmss}
\end{equation}

\subsection{Complex 3-Tensors}
\label{complexten}

Let us now consider the complex 3-tensors with $U(N)^3$ or $U(N)^2 \times O(N)$ symmetries. The latter symmetry is particularly interesting
because it is preserved by the tetrahedral interaction $\phi^{a_{1}b_{1}c_{1}}\bar \phi^{a_{1}b_{2}c_{2}}\phi^{a_{2}b_{1}c_{2}} \bar \phi^{a_{2}b_{2}c_{1}}$.
This means that there are interacting melonic theories with the $U(N)^2 \times O(N)$ symmetry \cite{Tanasa:2011ur,Tanasa:2015uhr,Klebanov:2016xxf}.

In the $U(N)^3$ case we have the fields $\phi^{abc }$ and $\bar{\phi}^{abc}$, which are in the tri-fundamental representations 
$N \times N \times N$ and
$\bar{N} \times \bar{N} \times \bar{N}$ respectively. The partition function reads:
\be
\mathcal{Z}^{U(N)^3} = \int \ dM_1 dM_2 dM_3 \exp \bigg( \sum_{m=1}^{\infty} \cfrac{z(x^m)}{m} \left( 
\chi(M^m_1) \chi(M^m_2) \chi(M^m_3) + 
\bchi(M^m_1) \bchi(M^m_2)
\bchi(M^m_3) \right)  \bigg)\,.
\ee
It is straightforward to compute it using the following large $N$ result\cite{Beccaria:2017aqc}:
\be
\label{U_calc}
\int dM \prod_{l \ge 1} (\tr M^l)^{a_l} (\tr \bar{M}^l)^{b_l}  = \prod_{l \ge 1} l^{a_l} a_l! \delta_{a_l, b_l}\,.
\ee
For the scalar we take $z_{S,0}(x)=x$ and find
\be
\mathcal{Z}^{U(N)^3}=1+ x^2+4x^4+11x^6+43x^8+161x^{10}+\dots\,.
\ee 
This expansion matches the results obtained in \cite{deMelloKoch:2017bvv} using group-theoretic methods.
 Extracting from $\mathcal{Z}$ the single-sum partition function, we find
\be
 \mathcal Z_{\text{s.s.}}^{U(N)^3}=x^2+3 x^4+7 x^6+26 x^8+97 x^{10}+\dots\,.
\ee
The coefficient $3$ of $x^4$ is in agreement with the fact that the tetrahedron invariant is not allowed by
the $U(N)^3$ symmetry. Only the 3 pillow invariants are allowed, and their form is
\be
\phi^{a_{1}b_{1}c_{1}}\bar \phi^{a_{1}b_{1}c_{2}}\phi^{a_{2}b_{2}c_{2}} \bar \phi^{a_{2}b_{2}c_{1}}\ ,
\qquad
\phi^{a_{1}b_{1}c_{1}}\bar \phi^{a_{1}b_{2}c_{1}}\phi^{a_{2}b_{2}c_{2}} \bar \phi^{a_{2}b_{1}c_{2}}\ ,
\qquad
\phi^{a_{1}b_{1}c_{1}}\bar \phi^{a_{2}b_{1}c_{1}}\phi^{a_{2}b_{2}c_{2}} \bar \phi^{a_{1}b_{2}c_{2}}\ .
\ee
The asymptotic number of operators can be estimated as follows. As in the $O(N)$ case, the integral
(\ref{U_calc}) grows factorially in $a_l$ and only as a power in $l$. It means that the term with $m=1$ again dominates.
Besides, to get a non-zero answer we need to extract the term with an equal number of $\chi(M_i)$ and $\bchi(M_i)$.
Therefore,
\beq
n^{U(N)^3}_{2k} \sim { {2k}\choose{k}  } \cfrac{1}{(2k)!} \int dM_1 dM_2 dM_3 \ \prod_{i=1}^3 \chi(M_i)^k \bchi(M_i)^k \sim k!
\eeq

In the $U(N)^2 \times O(N)$ case we have representations $N \times N \times N$ and $\bar{N} \times \bar{N} \times N$, so that
\beq
\mathcal{Z}^{U(N)^2 \times O(N)} = \int  dM_1 dM_2 dM_{3} \exp \bigg( \sum_{m=1}^{\infty} \cfrac{z(x^m)}{m} \big( \chi(M^m_1) \chi(M^m_2)+  \bchi(M^m_1)  
\bchi(M^m_2)\big)\chi(M_{3}^m)  \bigg)\,,
\eeq
where the matrices $M_1,M_2$ belong to $U(N)$, while $M_3$ belongs to $O(N)$.
The scalar partition function has the following expansion:
\be
\mathcal{Z}^{U(N)^2 \times O(N)}=1+x^2+6x^4+21x^6+147x^8+1043x^{10}+\dots\,.
\ee
Extracting the single-sum partition function, we find
\be
 \mathcal Z_{\text{s.s.}}^{U(N)^2 \times O(N)}=x^2+5x^4+15 x^6+111 x^8+821 x^{10}+\dots\,.
\ee
The coefficient $5$ of $x^4$ is in agreement with the fact that, addition to the tetrahedron invariant, there are 4 pillow invariants allowed by the
$U(N)^2 \times O(N)$ symmetry:
\begin{align}
& \phi^{a_{1}b_{1}c_{1}}\bar \phi^{a_{1}b_{1}c_{2}}\phi^{a_{2}b_{2}c_{1}} \bar \phi^{a_{2}b_{2}c_{2}}\ ,
\qquad
\phi^{a_{1}b_{1}c_{1}}\bar \phi^{a_{1}b_{1}c_{2}}\phi^{a_{2}b_{2}c_{2}} \bar \phi^{a_{2}b_{2}c_{1}}\ , \notag \\
& \phi^{a_{1}b_{1}c_{1}}\bar \phi^{a_{1}b_{2}c_{1}}\phi^{a_{2}b_{2}c_{2}} \bar \phi^{a_{2}b_{1}c_{2}}\ ,
\qquad
\phi^{a_{1}b_{1}c_{1}}\bar \phi^{a_{2}b_{1}c_{1}}\phi^{a_{2}b_{2}c_{2}} \bar \phi^{a_{1}b_{2}c_{2}}\ .
\end{align}
Using the same method as in the $U(N)^3$ case, the asymptotic growth can be found to be
\be
n^{U(N)^2 \times O(N)}_{2k} \sim 2^k k!
\ee

\section{The Hagedorn Transition}
\label{sec:Hagedorn}
The special features of the thermodynamics of free theories where the fields are tensors of rank $r\geq 3$ under some global symmetry group were
recently studied in  \cite{Beccaria:2017aqc}. It was found that the Hagedorn temperature vanishes in the large $N$ limit as $\sim 1/\log N$ \cite{Beccaria:2017aqc}.
In this section we show that this also applies to the models with $O(N)^3$ symmetry studied in this paper.

An essential feature of the large $N$ tensor models is that
the low temperature expansion of the partition function has the approximate structure $\sum_{k} 2^k k! x^{2k}$, where $-\ln x$ is proportional to $\beta$.
This power series is divergent and non-Borel summable; therefore, strictly speaking the partition function
is not defined for any finite temperature. 
To illustrate the basic points, we study the large $N$ behavior of the 
integral (\ref{Z_S_0}) in a standard fashion (it will be convenient to assume that $N$ is even).
First of all, for large $N$ there should be no difference between $SO(N)$ and $O(N)$. An $SO(N)$ matrix can always be put in the block-diagonal form with $2\times 2$ blocks corresponding to a rotation 
by an angle $\alpha^i$ in 2d plane. Including the $SO(N)$ measure \cite{Krauth:2000bv}, the partition function 
(\ref{Z_S_0}) can be rewritten as:
\begin{equation}
\mathcal{Z}=\int \prod_{r=1}^3 d \al^i_r \prod_{i < j}^{N/2} \sin^2{\cfrac{\al_r^i - \al^j_r}{2}} 
\sin^2{\cfrac{\al_r^i + \al^j_r}{2}} 
\exp \bigg( 8 \sum_{m=1}^{\infty} \cfrac{z(x^m)}{m} \prod_{r=1}^{3} \sum_{i=1}^{N/2} \cos(m \al^i_r)   \bigg) = \int [d \al] e^{-S_\text{eff}}\,.
\label{}
\end{equation}
Index $r$ labels different $SO(N)_{r}$ groups and $i,j=1,\dots,N/2$ go over rotation angles. Also we have introduced
a single-letter partition function $z(x)$ to work in more generality. The above equation is valid for
scalars, while for fermions we need to include the factor $(-1)^{m+1}$ in front of $z(x^m)$. 
However, we will see in a moment that for the Hagedorn
transition only $m=1$ term is relevant. Therefore, our main results will be applicable for both cases. 

The effective action
$S_\text{eff}$ reads
\begin{equation}
S_\text{eff} = -  \cfrac{1}{2} \sum_{r=1}^{3} \sum_{i \neq j}^{N/2} \bigg( \log \sin^2{\frac{\al_r^i - \al^j_r}{2}} 
+ \log \sin^2{\frac{\al_r^i + \al^j_r}{2}} \bigg)  - 8 \sum_{m=1}^{\infty} \cfrac{z(x^m)}{m} \prod_{r=1}^{3} \sum_{i=1}^{N/2} \cos(m \al^i_r)\,.
\label{}
\end{equation}
There are three saddle-point equations. One of them is:
\be
\sum_{j=1}^{N/2} \bigg( \cot{\cfrac{\al_1^i - \al^j_1}{2}} + \cot{\cfrac{\al_1^i + \al^j_1}{2}} \bigg) -8 \sum_{m=1}^{\infty} z(x^m)
\sin(m \al^i_1) \sum_{j_2, j_3} \cos(m \al^{j_2}_2) \cos(m \al^{j_3}_3) =0 \,.
\ee
The other two can be obtained by cyclic permutations of $\al^i_1,\al^i_2,\al^i_3$.
Introducing density functions:
\be
\rho_r(\al) = \cfrac{2}{N} \sum_{i=1}^{N/2} \delta(\al-\al^i_r)\,.
\ee
The saddle-point equation can be rewritten as:
\be
\int_{-\pi}^{\pi} d\al_1' \rho_1(\al_1') \bigg( \cot{\cfrac{\al_1 - \al_1'}{2}} + \cot{\cfrac{\al_1 + \al_1'}{2}} \bigg) -
4N \sum_{m=1}^{\infty} z(x^m) \sin(m \al_1) \rho^m_2 \rho^m_3=0\,,
\ee
where 
\be
\rho^m_r = \int_{-\pi}^\pi d \al \rho_r(\al) \cos(m \al)\,.
\ee
It is natural to assume that because of the cyclic symmetry $\rho_1=\rho_2=\rho_3=\rho(\al)$. Moreover, we will assume that 
$\rho$ is an even function: $\rho(\al)=\rho(-\al)$. With these assumptions the saddle-point equation reads as:
\be
2 \int_{-\pi}^{\pi} d\al' \rho(\al') \cot{\cfrac{\al - \al'}{2}}  -
4N \sum_{m=1}^{\infty} z(x^m) \sin(m \al) ( \rho^m )^2 =0\,.
\ee
This is exactly the saddle-point equation studied in \cite{Beccaria:2017aqc}, with their $6N$ replaced
by our $2N$. They have found that there is Hagedorn transition: for low temperatures when $N z(x)<27/16$ the 
partition function is dominated by the uniform saddle 
\be
\rho(\al) = \cfrac{1}{2 \pi}\,, \quad  \al \in [-\pi,\pi] \,.
\ee
And so all $\rho^m$ are zero for $m>0$. For higher temperatures, the density
$\rho$ is not a constant and takes non-zero values only within a smaller interval $[-\al_0,\al_0]$. Moreover, 
the transition point itself can be found by assuming that only $\rho^1$ becomes non-zero. Therefore, the transition
takes place at $N z(x) = 27/16$ for both bosons and fermions as we have advertised above.
More details
can be found in \cite{Beccaria:2017aqc} and \cite{Aharony:2003sx}.

For example, we can study the fermions in $d=1+2\epsilon$. According to eq. (\ref{z_f_1+e}), in the UV the transition
happens at
\be
z_{F,1+2\epsilon} = x^\epsilon = \exp(-\beta \epsilon) = \frac{27}{16 N}\,.
\ee
In the IR the fermions have dimension $1/4$ for $d=1$. Assuming that most $2k-$fermion operators have
dimension $k/2$,  we conclude that the transition takes place at:
\be
z_{F,\textrm{IR}}=x^{1/4}=\exp(-\beta/4) = \frac{27}{16 N}\,. 
\ee

\section*{Acknowledgments}

We are very grateful to E. Witten for important input into many aspects of this paper.
We are also grateful to S. Minwalla for important discussions and for sharing a draft of the paper \cite{Choudhury:2017tax} prior to publication.
We also thank  I. Danilenko, A. Jevicki, C. Krishnan, J. Maldacena, C. Peng, F. Popov, D. Roberts, S. Shenker and D. Stanford for useful discussions.
The work of IRK and GT was supported in part by the US NSF under Grant No.~PHY-1620059. GT acknowledges the support of a Myhrvold-Havranek Innovative Thinking Fellowship.


\bibliographystyle{ssg}
\bibliography{Operators}

\end{document}